\documentclass[journal,10pt]{IEEEtran}

\usepackage{graphicx}
\usepackage[colorlinks,linkcolor=black,anchorcolor=black,citecolor=black]{hyperref}
\usepackage{lineno,hyperref}
\modulolinenumbers[5]
\usepackage[cmex10]{amsmath}
\usepackage{amsfonts}

\usepackage{color}
\usepackage{algorithm}
\usepackage{algorithmic}
\usepackage{array}
\usepackage{multirow}
\usepackage{booktabs}
\usepackage{subfigure}
\usepackage{bm}
\usepackage{graphicx}
\usepackage{amsthm}
\usepackage{cite}
\usepackage{stfloats}

\allowdisplaybreaks[4]
\usepackage{amsmath}
\theoremstyle{definition}
\newtheorem{definition}{Definition}
\usepackage{underscore}
\usepackage{stfloats}
\usepackage{float}
\usepackage{lipsum}

\newtheorem{example}{Example}

\makeatletter
\def\subsubsection{\@startsection{subsubsection}{3}{\parindent}{0.5ex plus 1ex minus 0.1ex}{0pt}{\normalfont\normalsize}}

\begin{document}
\title{Knowledge-Driven 3D Semantic Spectrum Map: KE-VQ-Transformer Based UAV Semantic Communication and Map Completion}

\author{
Wei Wu,~\IEEEmembership{Member,~IEEE,}
Lingyi Wang,
Fuhui Zhou,~\IEEEmembership{Senior Member,~IEEE,}\\
Zhaohui Yang,~\IEEEmembership{Senior Member,~IEEE,}
Qihui Wu,~\IEEEmembership{Fellow,~IEEE}

\thanks{Copyright (c) 2025 IEEE. Personal use of this material is permitted. However, permission to use this material for any other purposes must be obtained from the IEEE by sending a request to pubs-permissions@ieee.org.}
\thanks{
Wei Wu is with the College of Communication and Information Engineering,
Nanjing University of Posts and Telecommunications, Nanjing, 210003, China,
also with the Anhui Province Key Laboratory of Cyberspace Security Situation Awareness and Evaluation, Hefei, 230037, China,
and also with the National Mobile Communications Research Laboratory, Southeast University, Nanjing, 210096, China
(e-mail: weiwu@njupt.edu.cn).}
\thanks{Lingyi Wang is with the College of Science,
Nanjing University of Posts and Telecommunications, Nanjing, 210003, China
(e-mail: lingyiwang@njupt.edu.cn).}
\thanks{Fuhui Zhou and Qihui Wu are with the College of Electronic and Information Engineering, 
Nanjing University of Aeronautics and Astronautics, Nanjing, 210000, China
(e-mail: zhoufuhui@ieee.org, wuqihui2014@sina.com).}
\thanks{Zhaohui Yang is with Zhejiang Lab, Hangzhou 311121, China, also with
the College of Information Science and Electronic Engineering, Zhejiang
University, Hangzhou, Zhejiang 310027, China, and also with the Zhejiang Provincial Key Laboratory of Information Processing, Communication and Networking (IPCAN), Hangzhou, Zhejiang 310007, China 
(e-mail: yang_zhaohui@zju.edu.cn).}
}

\maketitle

\begin{abstract}
    Artificial intelligence (AI)-native three-dimensional (3D) spectrum maps are crucial in spectrum monitoring for intelligent communication networks.
    However, it is challenging to obtain and transmit 3D spectrum maps in a spectrum-efficient, computation-efficient, and AI-driven manner, especially under complex communication environments and sparse sampling data.
    In this paper, we consider practical air-to-ground semantic communications for spectrum map completion, 
	where the unmanned aerial vehicle (UAV) measures the spectrum at spatial points and extracts the spectrum semantics, 
	which are then utilized to complete spectrum maps at the ground device.
    Since statistical machine learning can easily be misled by superficial data correlations with the lack of interpretability,
    we propose a novel knowledge-enhanced semantic spectrum map completion framework with two expert knowledge-driven constraints from physical signal propagation models. 
    This framework can capture the real-world physics and avoid getting stuck in the mindset of superficial data distributions.
	Furthermore, a knowledge-enhanced vector-quantized Transformer (KE-VQ-Transformer) based multi-scale low-complex intelligent completion approach is proposed, 
    where the sparse window is applied to avoid ultra-large 3D attention computation, and the multi-scale design improves the completion performance.
    The knowledge-enhanced mean square error (KMSE) and root KMSE (RKMSE) are introduced as novel metrics for semantic spectrum map completion 
	that jointly consider the numerical precision and physical consistency with the signal propagation model,
    based on which a joint offline and online training method is developed with supervised and unsupervised knowledge loss.
    The simulation demonstrates that our proposed scheme outperforms the state-of-the-art benchmark schemes in terms of RKMSE.
\end{abstract}  
  

\IEEEpeerreviewmaketitle
\section{Introduction}
In the context of the Intelligent Internet of Everything (IoE) for 6G and beyond \cite{8869705,10622764,9475174}, 
the exponential growth of smart devices and ultra-scale connection exacerbate the issue of spectrum scarcity \cite{10186369,8755300,10257607}.
Spectrum maps, as a critical visual and knowledge representation tool, not only facilitate dynamic spectrum management \cite{8648450,9484687,10525070,11008499}, but also serve as a foundation for integrated sensing and communication (ISAC) and context-adaptive network optimization in future intelligent communication systems \cite{10904265,que2025cooperative,hu2025advancing,wang2025radiodiff}.
Specifically, a spectrum map visually represents the signal power distribution of spectrum points across different physical locations, 
enabling the identification of interference sources and the efficient exploitation of underutilized spectrum. Compared to two-dimensional (2D) spectrum maps \cite{li2021robust,10121582}, 
three-dimensional (3D) spectrum maps \cite{jie20243d} display the spatial complexity inherent in spectrum characteristics and offer enhanced spatial resolution, 
capturing the intricate interactions between air and ground communication networks \cite{9484687}. Despite the advantages and importance of 3D spectrum maps, constructing 3D spectrum maps presents significant challenges. 
First of all, it is difficult to accurately and completely capture spatial spectrum data. 
Although, unmanned aerial vehicles (UAVs) equipped with spectrum monitoring devices have emerged as a critical solution for spatial spectrum data collection due to mobility and flexibility, limitations such as time delays and energy consumption, constrain the ability of UAVs to collect large-scale spectrum data. 
Secondly, 3D spectrum map construction involves the transmission of large volumes of spectrum data. It is challenging to ensure efficient communication over physical channels, particularly under constrained communication resources.
Hence, communication networks for spectrum maps need both accurate completion and efficient transmission.

Semantic communication for spectrum map construction provides a promising solution for the critical challenges mentioned above \cite{10525070}.
Recently, task-oriented semantic communication has received widespread research attention as a promising technology since it can adaptively extract critical semantic information specific to a given task while discarding irrelevant data \cite{10554663,lu2021reinforcement,10122224,10445328,xin2023information,10639525},
thus significantly enhancing the spectrum efficiency and ensuring robust data transmission \cite{10419853,10032275,10422980,du2023ai}.
Deep learning (DL) methods play a central role in enabling task-oriented semantic communication by learning intrinsic patterns and contextual relationships within data to preserve the critical semantics \cite{weng2023deep,ma2023task,liu2023adaptable,10638143,10158994}. 
However, the process of extracting and processing semantic information, particularly in constructing 3D spectrum maps, requires substantial computational resources. 
This poses a challenge for real-time processing, especially when spectrum data is collected by mobile platforms such as UAVs in complex terrains. 
Hence, semantics can be transmitted to the ground devices equipped with more powerful computational resources over the air-to-ground channel for spectrum map completion.  
In this way, an artificial intelligence (AI)-native communication network is formed to simultaneously enhance the data transmission efficiency and perform the 3D spectrum map completion task.

\vspace{-0.3cm}
\subsection{Related Work}
\subsubsection{\textit{Spectrum Map Completion}}
Intelligent spectrum map completion methods based on DL have been widely utilized with the prediction ability \cite{10025551,10103465,10757328,9857777}.
However, the majority of the works \cite{10025551,10103465,10757328,9857777} considered the spectrum map completion on the local device 
ignoring the transmission problem of a large number of spectrum data. 
In other words, to the author's best knowledge, there has been no work that completely investigates the artificial intelligence (AI)-driven spectrum semantic extraction, transmission, and completion.
The generative adversarial network (GAN) based intelligent completion schemes were used in \cite{10025551,10103465,10757328}, 
which aimed to deal with the non-ideal data and utilize the unsupervised learning capability of GAN, 
considering that it was difficult to capture the complete information in a large space.
The authors in \cite{9857777} jointly used a Bayesian estimator and data-driven DL algorithm to actively select the UAV location.
However, it is high-cost for UAVs to perform local processing of ultra-scale spectrum data, 
which requires extra large energy consumption and powerful computation capability \cite{8693989,8877759}.
Moreover, it is worth noting that the majority of existing data-driven intelligent spectrum completion methods \cite{10025551,9857777} generally suffer from poor interpretability and weak robustness,
although the AI-driven methods show a significant advantage compared to the traditional mathematical methods, such as inverse distance weighting (IDW).
Few works consider the spectrum map reconstruction over the complex air-to-ground channels, which raises further requirements for completion accuracy, computation efficiency, and spectrum efficiency, 
especially in low signal-to-noise ratio (SNR) cases.

\subsubsection{\textit{Semantic Communication Networks}}
DL-driven semantic coding schemes for semantic communications have been widely investigated \cite{9830752,10101778,wang2024adaptive}.
The authors in \cite{9830752} further introduced the unified semantic coding framework for multimodal multi-user semantic communication. Considering the semantic noise, the adversarial learning method was introduced in \cite{10101778}. 
Moreover, the reconstruction task performance in \cite{10101778} demonstrated the superior generation capabilities of autoencoders with crippled data.
The authors in \cite{wang2024adaptive} investigated the adaptive semantic-bit quantization method.
However, the above-mentioned works \cite{9830752,10101778,wang2024adaptive} solely applied statistical machine learning-driven semantic communication paradigm aims to approximate the optimal data distribution of the specific datasets,
vulnerable to the misrepresentation of superficial data correlations, 
particularly in the presence of strong interference or background noise.
This challenge makes it hard work to realize semantic communication and spectrum map completion over the complex air-to-ground wireless channels.

Some research attention has been paid to the knowledge-enhanced semantic communication \cite{10016636,zhou2023cognitive,10272264,10122227,wheeler2023knowledgedriven,9928407},
which can provide semantic interpretability and reliability.
The knowledge graph was used in \cite{zhou2023cognitive} to obtain semantics in an intuitive and interpretable way.
Further extending to system-level tasks, the works in \cite{10016636} and \cite{10272264} investigated the causal reasoning enhanced semantic communication networks.
Besides, the authors in \cite{10122227} proposed a knowledge-enhanced semantic decoder, leveraging facts in the knowledge base to facilitate semantic reasoning and decoding processes.
Moreover, the authors in \cite{wheeler2023knowledgedriven} proposed the conceptual space with knowledge representation, which can describe the semantics in an intuitive and interpretable way. 
However, the works in \cite{10122227} and \cite{wheeler2023knowledgedriven} considered the full-constellation-based semantic coding paradigm \cite{10101778}, which is difficult to be applied in the digital communication systems.
A wireless semantic communication network with domain knowledge was proposed in \cite{9928407}.
Different from \cite{10122227} focused on unilateral knowledge reasoning at the receiver,
the authors in \cite{9928407} introduced a dual-path framework with joint features and knowledge process at the both transmitter and receiver.
It is worth noting that the works \cite{10016636,10272264,zhou2023cognitive,10122227,wheeler2023knowledgedriven,9928407} consider a framework-level embedding of knowledge, 
which characterizes a dependence on enhanced algorithmic frameworks or additional technologies.
Since the spectrum completion follows the real-world physical model, we can explore an AI-native semantic communication network for the robust and interpretable spectrum map completion.

\vspace{-0.3cm}
\subsection{Contributions}
In this paper, a novel knowledge-enhanced semantic communication framework for 3D spectrum map completion is proposed to avoid the semantic representation stuck in the misrepresentation of superficial data correlations while enhancing the robustness, efficiency and interpretability of spectrum map completion. 
In the proposed framework, the real-world physical knowledge of signal propagation and channel fading are firstly utilized to constrain the construction of the semantic spectrum map. 
The main contributions of this paper are summarized as follows.
\begin{itemize}
  \item This work is the first to propose a semantic communication network with joint spectrum data transmission and 3D spectrum map completion, characterized by its robustness, transmission efficiency and AI-native design. 
  A practical air-to-ground scenario is considered, 
  where the UAV conducts the sparse measurement of signal power, extracts spectrum semantics, and transmits the spectrum semantics over the physical channel, and the ground user completes the spectrum map,
  with limited spectrum data based on the spectrum semantics.
  
  \item We propose a novel knowledge-enhanced semantic framework for spectrum map completion. 
  Specifically, we extract two prior knowledge from the physical free space signal propagation model as the constraints on spectrum map construction. 
  Based on two knowledge constraints, we further propose a knowledge-driven joint online and offline training method, 
  which aims to enable the semantic communication network to understand the real-world physical model, 
  thereby enhancing the interpretability and robust performance of the semantic communication network. 
  Furthermore, we define the knowledge-enhanced mean square error (KMSE) that considers the knowledge consistency along with mean square error (MSE)
  which can better evaluate the construction performance of the spectrum map at the semantic level.
  
  \item Based on the proposed framework, we further present a multi-scale low-complexity implementation scheme called knowledge-enhanced vector-quantized Transformer (KE-VQ-Transformer),
  where the sparse window avoids ultra-large 3D attention computation, and the multi-scale design improves the robustness of the spectrum map completion.

  \item The simulation demonstrates the superior completion performance of the proposed KE-VQ-Transformer compared to the state-of-the-art benchmark schemes in terms of the root KMSE (RKMSE).
  Moreover, the spectrum map constructed by the proposed knowledge-driven scheme can very closely approximate the original spectrum map generated by the free space signal propagation model, even at low SNRs or low sampling ratios. 
\end{itemize}

\vspace{-0.3cm}
\subsection{Organization And Notation}
The remainder of this paper is organized as follows. 
Section II presents the 3D free space signal propagation model and UAV semantic communication model. 
Section III presents the formulated semantic spectrum completion problem.
In Section IV, our proposed knowledge-enhanced semantic spectrum completion framework is presented.
Then, Section V introduces the KE-VQ-Transformer-based semantic communication network for spectrum map completion and the two-stage training method.
Section VI presents the simulation results.
Finally, Section VII concludes this paper.

The real matrix set with size $x \times y$ is denoted by $\mathbb{R}^{x \times y}$,
and the element set is denoted by $\{\cdot\}$. 
The vector, matrix, and the transpose of the vector are respectively denoted by $\boldsymbol{v}$, $\boldsymbol{M}$ and $\boldsymbol{v}^T$. 
The function $\arg \min_x g(x)$ denotes the value of $x$ that can minimize $g(x)$. 
The $(\cdot) \sim \mathcal{N}\left(\mu, \sigma^2\right)$ denotes $(\cdot)$ follows a normal distribution with mean $\mu$ and variance $\sigma$.

\section{System Model and Problem Formulation}
\subsection{3D-Free Space Signal Propagation Model}\label{2a}
\begin{figure*}
	\centering
	\includegraphics[scale=0.75]{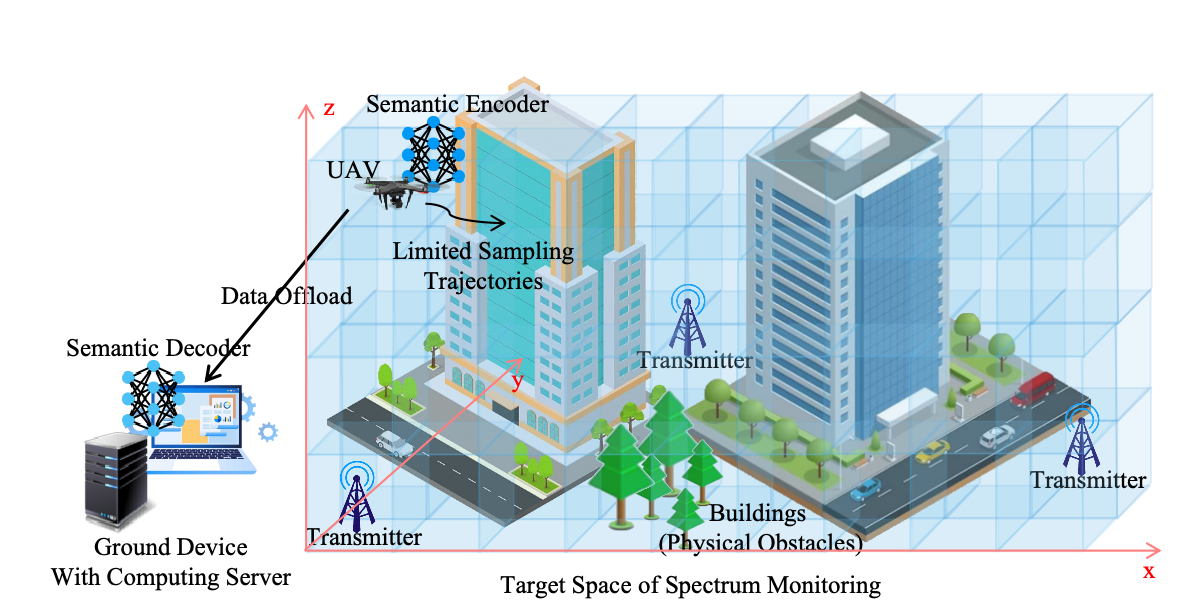}
	\caption{A showcase of UAV-enabled semantic communications for 3D spectrum monitoring under limited sampling trajectories.}
	\vspace{-0.4cm}
\end{figure*}
As illustrated in Fig.~1, we model a 3D spectrum monitoring space $\boldsymbol{S}$ within a Cartesian coordinate system. 
To better facilitate the measurement and characterize the coordinate positions, the target space $\boldsymbol{S}$ is first considered as a cuboid with the dimensions of $L \times W \times H$, 
where $L$, $W$ and $H$ are respectively the length, width, and height of $\boldsymbol{S}$.
Then the space $\boldsymbol{S}$ is divided into $N_{\mathrm{B}} = N_{\mathrm{L}} \times N_{\mathrm{W}} \times N_{\mathrm{H}}$ uniform blocks, 
where $N_{\mathrm{L}}$, $N_{\mathrm{W}}$ and $ N_{\mathrm{H}}$ respectively represent correspond to the number of blocks along the length, width, and height.
To ensure spatial separation and mitigate interference, a minimum separation between transmitters, i.e., at most one transmitter is located within each block, is assumed if multiple independent one-antenna transmitters exist.
Let $\mathcal{X}$ represent the set of block center coordinates, where each point $x\in\mathcal{X}$ is a discrete position in $\mathbb{R}^3$.
Moreover, there are $n$ active transmitters in $\boldsymbol{S}$, $n \in \{1,2,\cdots,N_T\}$, where $N_T$ represents the maximum possible number of transmitters, while the exact number $n$ is unknown for the UAV collector.
For the $i$-th active transmitter, $i \leq n$, its block location is represented by $x_{\mathrm{T},i} \in \mathcal{X}$, and its transmitter power at a given frequency $f$ is expressed as $\Omega_{f}(x_{\mathrm{T},i})$.
It is worth noting that the frequency $f$ can be extended to represent any point of interest in the spectrum.

Similar to \cite{10025551}, a signal propagation model that follows the log-distance path loss is considered.
The path loss $\mathrm{PL}(d_{x_{\mathrm{T},i}, x})(\mathrm{dB})$ from the location $x_{\mathrm{T},i}$ to $x$ is calculated by 
\begin{equation}\label{knowledge}
	\mathrm{PL}(d_{x_{\mathrm{T},i}, x})= -10 \log _{10} \frac{G_t \lambda^2}{(4 \pi d_{x_{\mathrm{T},i}, x})^2} +G_{\theta},
\end{equation}
where $d_{x_{\mathrm{T},i}, x}$ is the distance between the transmitter $x_{\mathrm{T}, i}$ and the receiver $x$, $G_t$ represents the antenna gain, 
the term $\lambda$ is the wavelength of the signal, and the term $G_\theta \sim \mathcal{N}\left(0, \theta^2\right)$ characterizes the noise fading with unit of $\mathrm{dB}$. 
Here, we use the operator $\phi_f(x;x_{T,i}) = 10^{-\mathrm{PL}_f(d_{x_{T,i},x})/10}$ to function as the signal propagation from the transmitter $x_{\mathrm{T}, i}$ to the receiver $x$.
We consider a $N_T$-transmitter scenario.
The received signal power at the location $x$ is expressed by 
\begin{equation}
	\Omega_{f}(x)=\sum_{i=1}^{N_T} \phi_f(x;x_{\mathrm{T}, i}) \Omega_{f}\left(x_{\mathrm{T}, i}\right)+A_\sigma,
\end{equation}
where the term $A_\sigma$ is the additive white Gaussian noise (AWGN), $A_\sigma \sim \mathcal{N}\left(0, \sigma^2\right)$.
Hence, the actual 3D spectrum map is represented by $\boldsymbol{\Omega}_f=\{\Omega_f(x) \mid \forall x \in \boldsymbol{A}\}$,
where $\boldsymbol{A}$ represents the point set of the spectrum space.

However, in practical applications, the formidable challenges posed by the measurement loss and spatial constraints constrain our ability to sample substantial data points. 
Hence, we can only obtain an incomplete 3D spectrum map $\widetilde{\boldsymbol{\Omega}}_f=\{\widetilde{\Omega}_f(x) \mid \forall x \in \boldsymbol{A}\}$ from the UAV collected data.
Considering practical geographic constraints and flight safety requirements, the trajectory of the UAV is confined to a predefined admissible path, based on which we have $N_j$ measured points.
Let $\widetilde{\Omega}_f\left(x_{\mathrm{M}, j}\right)$ be the signal power at the $j$-th measurement point, $\widetilde{\Omega}_f\left(x_{\mathrm{M}, j}\right)=\Omega_f\left(x_{\mathrm{M}, j}\right)$.
Moreover, the unknown points are required to be completed, represented by $\{\widetilde{\Omega}_f\left(x_{\mathrm{U}}\right)\}$, $x_{\mathrm{U}} \in \{ \boldsymbol{A} \backslash \{x_{\mathrm{M}, j}\} \}$.
Taking into account the subsequent spectrum control and monitoring, 
the spectrum map is transmitted from the UAV to the ground user.
Notably, the intricate process of inferring and complementing the 3D spectrum map is conducted on the ground due to the energy and computational power limitations of UAVs.
Let $\widehat{\boldsymbol{\Omega}}_f=\{\widehat{\Omega}_f(x) \mid \forall x \in \boldsymbol{A}\}$ be the reconstructed spectrum map,
and our goal aims to render the $\widehat{\boldsymbol{\Omega}}_f$ as analogous to $\boldsymbol{\Omega}_f$ as feasibly achievable.

\subsection{UAV Semantic Communication Model}
\vspace{-0.1cm}
We introduce the concept of semantic communication for the transmission of spectrum data, driven by two fundamental considerations. 
Firstly, we aim to reduce transmission overhead by selectively extracting and transmitting semantic features derived from the high-dimensional spectrum maps. 
Secondly, we aim to enhance the semantic reasoning and the completion of 3D spectrum maps at the ground device based on the received features.

The UAV, equipped with $M$ antennas and a semantic encoder, can measure the signal power in the continuous, sparse and limited trajectories due to physical constraints, such as flight altitude and building obstructions. During the training stage, the transmitter locations are assumed to be known to the UAV.
Due to the large amount of invalid data in $\boldsymbol{\Omega}_f$, i.e., unmeasured points, the semantic encoder aims to extract the semantic features 
and describe the semantic information $\boldsymbol{d}$ with spatial features.
Let $E_{\alpha}(\cdot)$ be the semantic encoder network with parameters $\alpha$. Given the input spectrum map $\boldsymbol{\Omega}_f$, the semantics $\boldsymbol{d}$ are extracted by $\boldsymbol{d}=E_{\alpha}(\boldsymbol{\Omega}_f)$ to capture the critical spectral structures, 
propagation characteristics, and spatial-frequency correlations relevant to the radio environment.
Unlike conventional signal features, $\boldsymbol{d}$ is not directly interpretable due to the underlying physical dependencies and transmission distortions across the physical layer.
Hence, it is necessary to establish a common \textit{knowledge base} that enables both communicating parties to encode and decode $\boldsymbol{d}$ consistently in the semantic domain.

To this end, we consider a shared codebook-based semantic symbol set $\mathcal{D}=\{\widetilde{\boldsymbol{d}}_l\}_{l=0}^{L}$ for wireless semantic communications.
In this context, the semantic symbol set serves dual functions. Firstly, it acts as the fundamental \textit{semantic knowledge base} jointly known by the transmitter and receiver, 
and provides a set of \textit{discrete semantic symbols} that carry domain-specific knowledge of the radio environment. 
Second, the discrete indices of these codewords serve as compact symbolic representations of the continuous semantic features, 
thereby facilitating efficient transmission at the physical layer \cite{10622764}.
Hence, the UAV can quantify the semantics $\boldsymbol{d}$ by searching the most similar semantic symbol $\widetilde{\boldsymbol{d}}$ in the codebook.
Similar to \cite{10101778}, the minimum space distance is used to access the similarity between the semantic information and descriptions, 
represented by 
\vspace{-0.2cm}
\begin{equation}\label{ns}
	\widetilde{\boldsymbol{d}} =\arg \min _{\widetilde{\boldsymbol{d}}_{l} \in \mathcal{D}}\left\| \boldsymbol{d} - \widetilde{\boldsymbol{d}}_{l}\right\|_2.
\end{equation}
The index of the most similar description $\widetilde{\boldsymbol{d}}$ is considered as the representation for critical semantic information, and then transmitted over the physical channel.
It is worth noting that such a description set $\mathcal{D}$ not only needs to efficiently describe the semantic information $\boldsymbol{d}$ extracted from the incomplete map $\widetilde{\boldsymbol{\Omega}}_f$, 
but also allows the receiver to understand the semantic symbols $\widetilde{\boldsymbol{d}}$ and provides sufficient knowledge for 3D map complementation $\widehat{\boldsymbol{\Omega}}_f$.

The ground user, equipped with antennas and a semantic decoder, can interpret the received semantic information and utilize the semantic knowledge to complete the 3D spectrum map. The received baseband signal at the user is modeled as
$
y = \boldsymbol{h}x + A_m,
$
where $\boldsymbol{h}$ represents the complex channel coefficient between the UAV and the ground user, and $A_m$ represents the additive white Gaussian noise (AWGN) with $A_m \sim \mathcal{N}(0,m^2)$. 
During semantic transmission, the UAV transmits the quantized semantic symbol $\widetilde{\boldsymbol{d}}$ selected from the shared codebook. 
Due to the effect of the wireless channel, the received signal $y$ is distorted by both the fading term $\boldsymbol{h}$ and the noise term $A_m$, which can cause a symbol-level perturbation when decoding the index of $\widetilde{\boldsymbol{d}}$. 
Hence, the recovered semantic symbol $\widehat{\boldsymbol{d}}$ obtained at the ground user can not be identical to the transmitted $\widetilde{\boldsymbol{d}}$, i.e., $\widehat{\boldsymbol{d}} \neq \widetilde{\boldsymbol{d}}$. 
This mismatch reflects the semantic degradation introduced by the physical channel, where stronger fading or noise, i.e., smaller $|\boldsymbol{h}|$ or larger $m^2$, increases the probability of symbol distortion. 
Let $D_{\beta}(\cdot)$ denote the semantic decoder parameterized by $\beta$, then the completed semantic spectrum map is reconstructed as
$
\widehat{\boldsymbol{\Omega}}_f = D_{\beta}(\widehat{\boldsymbol{d}})
$.

\vspace{-0.1cm}
\section{Problem Formulation}
As considered in Section \ref{2a}, 
the target is to complete the unmeasured spectrum points accurately and reconstruct the entire 3D spectrum map. 
Hence, we expect a small gap between the actual spectrum map $\boldsymbol{\Omega}_f$ and the reconstructed spectrum map $\widehat{\boldsymbol{\Omega}}_f$.
In \cite{10025551}, the MSE loss function is used to evaluate this gap. 
Similarly, the 3D spectrum map reconstruction problem over the map set $\{\mathcal{X}\}$ is formulated as
\begin{subequations}\label{problem}
\begin{align}
&\mathbf{P}: \min _{\alpha,\beta,\mathcal{D}} \mathbb{E}_{\{\mathcal{X}\}}\left[\mathrm{MSE}(\Omega_f(x)-\widehat{\Omega}_f(x))\right]  \\
\text { s.t. } & \mathrm{C1}: |\widehat{\Omega}_f\left(x_{\mathrm{M},j}\right) - \Omega_f\left(x_{\mathrm{M}, j}\right)|< \epsilon, \forall j \in\{1,2,\cdots,J\},
\end{align}
\end{subequations}
where the averaged experiment loss $\mathrm{MSE}(\Omega_f(x)-\widehat{\Omega}_f(x)) = \frac{1}{N_{\mathrm{L}} \times N_{\mathrm{W}} \times N_{\mathrm{H}}} \sum_{x \in \mathrm{A}}(\Omega_f(x)-\widehat{\Omega}_f(x))^2$, $x \in \mathcal{X}$, 
and $J$ represents the limited number of measurements, $J \ll N_{\mathrm{B}}$.
Constraint $\mathrm{C1}$ aims to reduce the estimated errors at the location $x_{\mathrm{M},j}$, where $\epsilon$ is a positive value close to zero.

Traditional completion metrics, such as MSE, are limited to data-level evaluations. 
A smaller data deviation does not necessarily correspond to better adherence to physical rules, making such metrics insufficient for semantic communications to characterize physically consistent outcomes.
The fundamental challenge lies in the inability of semantic coding to explicitly learn the laws of physics when relying solely on MSE. 
This limitation not only limits the coding model to capture physics but also results in low interpretability and reduced stability of the semantic outputs.
To solve the above-mentioned problem, we propose a novel KMSE metric (12) in Section \ref{sknow} that jointly considers the MSE from the spectrum data and the physical deviation from the physical free space signal propagation model, which can impose knowledge-driven guidance on the semantics.
Hence, the 3D spectrum map reconstruction problem can be rewritten as
\vspace{-0.1cm}
\begin{subequations}\label{problem2}
	\begin{align}
	&\mathbf{P}^{\prime}: \min _{\alpha,\beta,\mathcal{D}} \mathbb{E}_{\{\mathcal{X}\}}\left[\mathrm{KMSE}(\Omega_f(x)-\widehat{\Omega}_f(x))\right] \\
	\text { s.t. } & \mathrm{C1}.
	\end{align}
\end{subequations}

\vspace{-0.5cm}
\section{Proposed Knowledge-Driven Semantic Spectrum Map Completion}\label{sknow}
In this section, we present a novel knowledge-driven semantic coding framework for spectrum map completion.
It is worth noting the traditional completion metric, MSE, remains on the data-level evaluation and fails to capture semantic-level completion performance, making it hard for semantic communications to characterize physically consistent outcomes by solely relying on the MSE.
Furthermore, existing semantic coding methods exhibit limited interpretability and stability, rendering them inadequate for the spectrum map completion task that requires strict adherence to underlying the physical signal propagation model.
Hence, we first extract two pieces of expert knowledge from the physical free space signal propagation model,
then we propose a novel KMSE metric that jointly considers the MSE from the spectrum data and the physical deviation, 
along with a joint online and offline training method based on knowledge-enhanced supervised learning and unsupervised learning.

\vspace{-0.3cm}
\subsection{Prior Expert Knowledge}\label{kf}
Note that the real-world physical model can enhance the interpretability and robust performance of the 3D semantic spectrum completion,
especially when the number of transmitters at the target monitoring space is unknown.
Hence, we adopt the spectrum knowledge to enhance the semantic coding network with interpretability and robustness.
Specifically, the free space signal propagation model is considered as a real-world physical model, and the 3D spectrum completion should follow this model.
The physical signal propagation model that characterizes is expressed as (\ref{knowledge}),
from which we can conclude two necessary expert prior knowledge which serves as knowledge-driven constraints as follows.

\begin{enumerate}
	\item Prior Knowledge 1: (Signal Propagation Direction Constraint) The signal propagation satisfies $\phi_f(x_1;x_{\mathrm{T}, i}) > \phi_f(x_2;x_{\mathrm{T}, i}), \forall d_{x_{\mathrm{T},i}, x_1} < d_{x_{\mathrm{T},i}, x_2}$.
  
	\item Prior Knowledge 2: (Signal Propagation Loss Constraint) The signal power satisfies $\widehat{\Omega}_f(x_{\mathrm{T},i}) - \widehat{\Omega}_f(x) = \Omega_f(x_{\mathrm{T},i}) - \Omega_f(x)$ when the estimation error is free, i.e., $\widehat{\Omega}_f(x_{\mathrm{T},i})=\Omega_f(x_{\mathrm{T},i})$ and $\widehat{\Omega}_f(x) = \Omega_f(x)$.
\end{enumerate}
From a physical perspective, Prior Knowledge 2 can be considered as the knowledge of signal fading during propagation. Note that Prior Knowledge 2 can be further simplified to $\widehat{\Omega}_f(x_1) - \widehat{\Omega}_f(x_2) = \Omega_f(x_1) - \Omega_f(x_2)$, which still works when the number and locations of transmitters are unknown.
Although the free-space path-loss model is adopted in this paper for clarity, the proposed knowledge constraints are model-agnostic and can incorporate any propagation formulation, e.g., log-distance, COST-231, or ray-tracing-based models. Hence, the knowledge framework can flexibly adapt to both line-of-sight (LoS) and non-line-of-sight (NLoS) environments by embedding the corresponding spectrum knowledge into the learning process.
Based on these two expert knowledge, we design the knowledge-driven semantic coding networks to better approximate the real-world physical model and enhance the quality of map completion.
First, the physical knowledge is serialized and integrated into the process of forming the common semantic knowledge base $\{\widetilde{\boldsymbol{d}}_{l}\}$. 
Then, the output spectrum map of semantic coding networks is guided and constrained by knowledge. More training details are given as the next three sections.

\vspace{-0.3cm}
\subsection{Offline Knowledge-Driven Semantic Loss}
\begin{figure*}[ht!]
	\begin{equation}\label{offlinelossknow}
		\begin{aligned}
\mathcal{L}^{\rm{S}}_{K} = \sum_i^{N_{\mathrm{T}}} \frac{1}{N_{\mathrm{l}} \times N_{\mathrm{w}} \times N_{\mathrm{h}}} & \{\sum_{x \in \widetilde{\boldsymbol{B}}_i} [(\Omega_f(x_{\mathrm{T},i})-\Omega_f(x))-(\widehat{\Omega}_f(x_{\mathrm{T},i})-\widehat{\Omega}_f(x))]^2 \\
&+\sum_{x \in \{\boldsymbol{B}_i / \widetilde{\boldsymbol{B}}_i\}} [(\sum_i^{N_{\mathrm{T}}} \rho(x_{\mathrm{T},i},x) \Omega_f(x_{\mathrm{T},i})-\Omega_f(x)]-[\sum_i^{N_{\mathrm{T}}} \rho(x_{\mathrm{T},i},x) \widehat{\Omega}_f(x_{\mathrm{T},i})-\widehat{\Omega}_f(x))]^2\}.
		\end{aligned}
	\end{equation}
\hrulefill
\end{figure*}
The semantic loss function is composed of three key components: data loss, knowledge loss, and communication loss.

The data loss focuses on learning the distribution of the training spectrum data, represented by
\begin{equation}\label{lossd}
	\begin{aligned}
\mathcal{L}_{D} = &\frac{1}{N_{\mathrm{L}} \times N_{\mathrm{W}} \times N_{\mathrm{H}}} [\sum_{x_{\mathrm{M}} \in  \{x_{\mathrm{M}, j}\}}(\Omega_f(x_{\mathrm{M}})-\widehat{\Omega}_f(x_{\mathrm{M}}))^2\\
&+\kappa \sum_{x_{\mathrm{U}} \in \{ \boldsymbol{A} \backslash \{x_{\mathrm{M}, j}\} \}} (\widehat{\Omega}_f\left(x_{\mathrm{U}}\right) - \Omega_f\left(x_{\mathrm{U}}\right) )],
	\end{aligned}
\end{equation}
where $\kappa < 1$ is the confident factor.
Based on eq. (\ref{lossd}), the reconstruction accuracy of the measured data is prioritized and guaranteed.
It is crucial to highlight that data loss serves as the training criterion for both the encoder and decoder in data information recovery. 
However, relying solely on data loss is insufficient for fostering semantic communication. A sole focus on data loss makes the semantic communication network  a `black box', which lacks  the interpretability and the understanding of the real physical model, 
resulting in diminished robustness and suboptimal generalization performance. 
This limitation constrains the application of semantic spectrum completion in dynamic wireless communication networks and real spectrum data beyond the training dataset.

Then, we introduce the knowledge loss, where the real-world physical model is explored to enable knowledge-driven robust semantic communication.
Let $\boldsymbol{B}_i$ be the space of correlation regions with the transmitter $i$, where each $\boldsymbol{B}_i$ can be divided into $N_{\mathrm{l}} \times N_{\mathrm{w}} \times N_{\mathrm{h}}$ blocks.
Jointly considering the prior knowledge 1 and knowledge 2, the knowledge loss is expressed by
$
\mathcal{L}_{K} = \sum_i^{N_{\mathrm{T}}} \frac{1}{N_{\mathrm{l}} \times N_{\mathrm{w}} \times N_{\mathrm{h}}} \sum_{x \in \boldsymbol{B}_i} ([\Omega_f(x_{\mathrm{T},i})-\Omega_f(x)]-[\widehat{\Omega}_f(x_{\mathrm{T},i})-\widehat{\Omega}_f(x)])^2.
$
However, there are still two challenges for the knowledge loss $\mathcal{L}_{K}$. 
\begin{enumerate}
	\item Challenge 1: It is hard work to access the exact number and accurate location of transmitters in the practical environment.
  
	\item Challenge 2: The simple knowledge loss $\mathcal{L}_{K}$ does not generalize the spatial overlap of signals that occurs in the case of multiple transmitters.
\end{enumerate}

To tackle Challenge 1, we independently propose a supervised knowledge loss for offline training, 
shown as follows, and an unsupervised knowledge loss for online training in Section \ref{okl}, where it is unnecessary to know the number and location of transmitters in advance.
Regarding Challenge 2, we extend the generalized case of knowledge loss with multiple transmitters and overlapped interacting spectrum spaces.
The non-overlapping correlation regions and the overlapping correlation regions
are respectively represented by $\widetilde{\boldsymbol{B}}_i$ and $\boldsymbol{B}_i / \widetilde{\boldsymbol{B}}_i$.
The supervised knowledge learning loss is rewritten as eq. (\ref{offlinelossknow}),
where $\rho(x_{\mathrm{T},i},x) \in \{0,1\}$ represents whether the transmitter at the $x_{\mathrm{T},i}$ has an impact on $x$.
If the signal from $x_{\mathrm{T},i}$ is propagated $x$, $\rho(x_{\mathrm{T},i},x)  = 1$; otherwise, $\rho(x_{\mathrm{T},i},x)  = 0$.
The loss function (\ref{offlinelossknow}) simultaneously limits the magnitude and direction of the signal decay.
Different from the data loss, the knowledge loss serves as the criterion constraint for the output of the decoder, i.e., the reconstructed 3D spectrum map.

The common knowledge base based on the codebook determines the efficiency of semantic communication between the parties. 
Specifically, this knowledge base is a collection of real-world knowledge and dataset experience designed to optimize the discrete delivery of semantics extracted at the transmitter and the precise understanding of semantics at the receiver.
The communication loss can be represented by 
\begin{equation}
\mathcal{L}_C=\left\|\operatorname{tg}[\boldsymbol{d}]-\widetilde{\boldsymbol{d}}\right\|_2^2+\gamma\left\|\boldsymbol{d}-\operatorname{tg}[\widetilde{\boldsymbol{d}}]\right\|_2^2,
\end{equation}
where $\gamma$ is a given weight factor, and $\operatorname{tg}[\boldsymbol{d}]$ truncates the gradient of $\boldsymbol{d}$.
The first term of $\mathcal{L}_C$ is quantization loss, 
which is used as a training criterion for codebook training to find an efficient communication knowledge base shared by both communicating parties. 
The second term of $\mathcal{L}_C$ is inductive loss, which is used as a training criterion for the encoder, 
aiming to encourage the encoder to use descriptions $\{\widetilde{\boldsymbol{d}}_l\}$ in the codebook to characterize the semantic output $\boldsymbol{d}$.
The total loss function for the offline training of knowledge-driven semantic communication networks is represented by
\begin{equation}\label{Off}
	\mathcal{L}^{\rm{Off}} = \mathcal{L}_{D} + w_{K} \mathcal{L}^{\mathrm{S}}_{K} + w_{C} \mathcal{L}_C,
\end{equation}
where $w_{K}$ and $w_{C}$ are balance coefficients.

\vspace{-0.3cm}
\subsection{Online Knowledge-Driven Semantic Loss}\label{okl}
Due to the impact of different physical factors in complex and dynamic radio environments, such as dynamic sampling points caused by building occlusion and background noises with different strengths,
offline-trained semantic communication networks are hard to apply to real-world scenarios with the transmitter locations unknown.
Hence, further taking advantage of the knowledge-driven constraints illustrated in Section \ref{kf}, 
we propose a novel unsupervised knowledge loss for the online training of semantic spectrum completion.
Specifically, the local peak points of the signal strength are reverse searched to be transmitters, and the physical model constraints, i.e., Knowledge Constraints 1 and 2, are applied to the region surrounding the evaluated transmitters.
\begin{definition}
	The estimated transmitter set is denoted by $\widetilde{\mathrm{T}}$, in which the transmitter location $x_{\widetilde{\mathrm{T}},i}$ satisfies
	\begin{subequations}\label{onst}
		\begin{align}
		& \exists x_{\widetilde{\mathrm{T}},i} \\
		\text { s.t. } & \widehat{\Omega}_f(x_{\widetilde{\mathrm{T}},i}) > \widehat{\Omega}_f(x_{\widetilde{\mathrm{T}},i}^{\kappa}), \forall \kappa \le \varphi \\
		& |\widehat{\Omega}_f(x_{\widetilde{\mathrm{T}},i}) - \widehat{\Omega}_f(x_{\widetilde{\mathrm{T}},i}^{\kappa}) - \mathrm{PL}(d_{x_{\widetilde{\mathrm{T}},i}, x_{\widetilde{\mathrm{T}},i}^{\kappa}})|<\zeta, \forall \kappa \le \varphi,
		\end{align}
	\end{subequations}
where $x_{\widetilde{\mathrm{T}},i}^{\kappa}$ represents the neighboring location of the estimated transmitter $x_{\widetilde{\mathrm{T}},i}$ with distance $\kappa$,
$\zeta$ represents the tolerable generation error,
and the $\varphi$ is the assessment region size that represents the accuracy of the potential transmitters.
The constraints (\ref{onst}a) and (\ref{onst}b) provide the local peak of signal strength, which is assumed to be the potential transmitter.
\end{definition}

Based on the estimated transmitters $\widetilde{\mathrm{T}}$, the unsupervised knowledge loss for online training of semantic spectrum map completion is presented by (\ref{onlinelossknow}),
where $N_{\widetilde{\mathrm{T}}}$ represents the number of the transmitters in $\widetilde{\mathrm{T}}$.
Similar to supervised knowledge loss (\ref{offlinelossknow}), the unsupervised knowledge loss is extended to the practical scenario with multiple transmitters and overlapped interacting spectrum spaces.
The estimated non-overlapping correlation regions and the estimated overlapping correlation regions for online training
are respectively represented by $\widetilde{\boldsymbol{O}}_i$ and $\boldsymbol{O}_i / \widetilde{\boldsymbol{O}}_i$.
It is clear that the unsupervised knowledge loss can implement knowledge-enhanced error correction on spectrum map outcomes,
where there is no need to know the exact prior information of the transmitters, such as the location and the transmit power.
To minimize the unsupervised knowledge loss under the dynamically practical radio environment, 
the network parameters $\alpha^*$ and $\beta^*$ are further updated to $\mathring{\alpha}^*$ and $\mathring{\beta}^*$ based on the gradient method with (\ref{offlinelossknow}). We note that the proposed online knowledge-driven loss provides a self-adaptive mechanism that continuously updates the network parameters without requiring labeled data, enabling effective operation in dynamic spectrum environments, such as varying channel states, user mobility, and partial blockage.
By minimizing the discrepancy between real-time observations and the embedded knowledge-based propagation constraints, the model can automatically correct deviations and adjust its semantic representations in response to environmental dynamics.
Although this paper focuses on static scenarios for clarity, the proposed knowledge-enhanced framework naturally extends to time-varying spectrum environments through the joint effect of online learning and spectrum knowledge adaptation.
\begin{figure*}
	\begin{equation}\label{onlinelossknow}
		\begin{aligned}
			\mathcal{L}^{\rm{Onl}} = \mathcal{L}^{\rm{US}}_{K} = \sum_i^{N_{\widetilde{\mathrm{T}}}}  \frac{1}{N_{\mathrm{l}} \times N_{\mathrm{w}} \times N_{\mathrm{h}}} & \{\sum_{x \in \widetilde{\boldsymbol{O}}_i} [\widehat{\Omega}_f(x_{\widetilde{\mathrm{T}},i})- \widehat{\Omega}_f(x) - \mathrm{PL}(d_{x_{\widetilde{\mathrm{T}},i}, x})]^2 \\
			&+\sum_{x \in \{\boldsymbol{O}_i / \widetilde{\boldsymbol{O}}_i\}} [(\sum_i^{N_{\widetilde{\mathrm{T}}}} \rho(x_{\widetilde{\mathrm{T}},i},x) \widehat{\Omega}_f(x_{\widetilde{\mathrm{T}},i})-\widehat{\Omega}_f(x))-(\sum_i^{N_{\widetilde{\mathrm{T}}}}\rho(x_{\widetilde{\mathrm{T}},i},x)\mathrm{PL}(d_{x_{\widetilde{\mathrm{T}},i}, x}))]^2\}.
		\end{aligned}
	\end{equation}
\hrulefill
\end{figure*}

\vspace{-0.2cm}
\subsection{KMSE Metric For Knowledge-Driven Map Completion Performance}
Since a single MSE term only measures the numerical difference between the generated and target maps, 
it cannot characterize semantic and physical consistency at both the sensory and physical layers. 
This limitation makes it inadequate for evaluating the performance of the semantic spectrum map completion. 
To overcome this issue, we propose a novel \textit{knowledge-enhanced mean square error (KMSE)} metric 
that jointly considers data fidelity and knowledge consistency. 
The KMSE is defined as
\begin{equation}
	\mathrm{KMSE}(\Omega_f(x)-\widehat{\Omega}_f(x)) = \mathrm{MSE}(\Omega_f(x)-\widehat{\Omega}_f(x)) + \mathcal{L}^{\mathrm{S}}_{K},
\end{equation}
where $\mathcal{L}^{\mathrm{S}}_{K}$ denotes the knowledge-driven constraint loss defined in (7). 
This design follows the principle of multi-objective optimization \cite{raissi2019physics,meng2025physics,bischof2025multi,mazandarani2025perception}, 
where both data fitting and physical knowledge consistency are jointly optimized through a linear weighted formulation. 
The linear additive form is widely adopted in hybrid data-knowledge learning frameworks, 
such as physics-informed \cite{raissi2019physics,meng2025physics,bischof2025multi} and knowledge-guided \cite{mazandarani2025perception} neural networks, 
because it maintains the interpretability of MSE while embedding domain knowledge as a regularization term.
Hence, the proposed KMSE can be regarded as a composite semantic metric that captures both the reconstruction fidelity at the sensory layer and the physical interpretability of the generated spectrum map.

For completeness, the root form of KMSE is further expressed as
\begin{equation}
	\mathrm{RKMSE}(\Omega_f(x)-\widehat{\Omega}_f(x)) = \sqrt{\mathrm{MSE}(\Omega_f(x)-\widehat{\Omega}_f(x)) + \mathcal{L}^{\mathrm{S}}_{K}}.
\end{equation}
Compared with the conventional MSE, the proposed KMSE and RKMSE metrics can jointly reflect the 
data reconstruction loss, sensory-layer consistency, and physical-model deviation, 
thus providing a more comprehensive evaluation of the semantic spectrum map completion performance.

\begin{example}[Necessity of KMSE]
Consider one transmitter and two spatial points, $x_{\text{near}}$ and $x_{\text{far}}$, with $d(x_{\text{near}})<d(x_{\text{far}})$ so the true spectrum satisfies $\Omega_f(x_{\text{near}}) > \Omega_f(x_{\text{far}})$. Let the normalized ground truth be $\Omega_f^\star(x_{\text{near}})=1.0$ and $\Omega_f^\star(x_{\text{far}})=0.8$.
We take two reconstructions $\widehat{\Omega}_f^{(A)}: (1.6,\,0.2)$ and $\widehat{\Omega}_f^{(B)}: (0.4,\,1.4)$, with the same MSE as
\begin{equation}
\mathrm{MSE}\big(\widehat{\Omega}_f^{(A)},\Omega_f^\star\big)=\mathrm{MSE}\big(\widehat{\Omega}_f^{(B)},\Omega_f^\star\big)
=0.36.
\end{equation}
Hence, MSE cannot distinguish which reconstruction is better.
However, $\widehat{\Omega}_f^{(B)}$ violates Prior Knowledge 2. By the proposed knowledge-driven KMSE defined in (12), we have
\begin{equation}
\mathrm{KMSE}\big(\widehat{\Omega}_f^{(A)}\big)=0.36,\qquad
\mathrm{KMSE}\big(\widehat{\Omega}_f^{(B)}\big)=1.36.
\end{equation}
Hence, KMSE can measure physical consistency even when MSE is identical. This example captures the general situation in 3D maps that local MSE can be low while violating physical propagation rules, and KMSE resolves this ambiguity by embedding the knowledge constraint $L_K^S$ into the metric.
\end{example}

\section{KE-VQ-Transformer Based Semantic Spectrum Map Completion With Two Stage Training}
In this section, we present a KE-VQ-Transformer scheme that realizes the spectrum completion framework proposed in Section III.
Firstly, we proposed a low-complex and multi-scale 3D-Transformer as the backbone of the knowledge-enhanced semantic spectrum completion network, 
which can significantly reduce the computation complexity and enhance the map completion.
Then, a two-stage training of the KE-VQ-Transformer is given, as shown in Fig. 2. 

\begin{figure*}
	\centering
	\subfigure[Stage 1: The training of the KE-VQ-Transformer based semantic coding network.]{
	  \begin{minipage}{14cm}
	  \includegraphics[width=\textwidth]{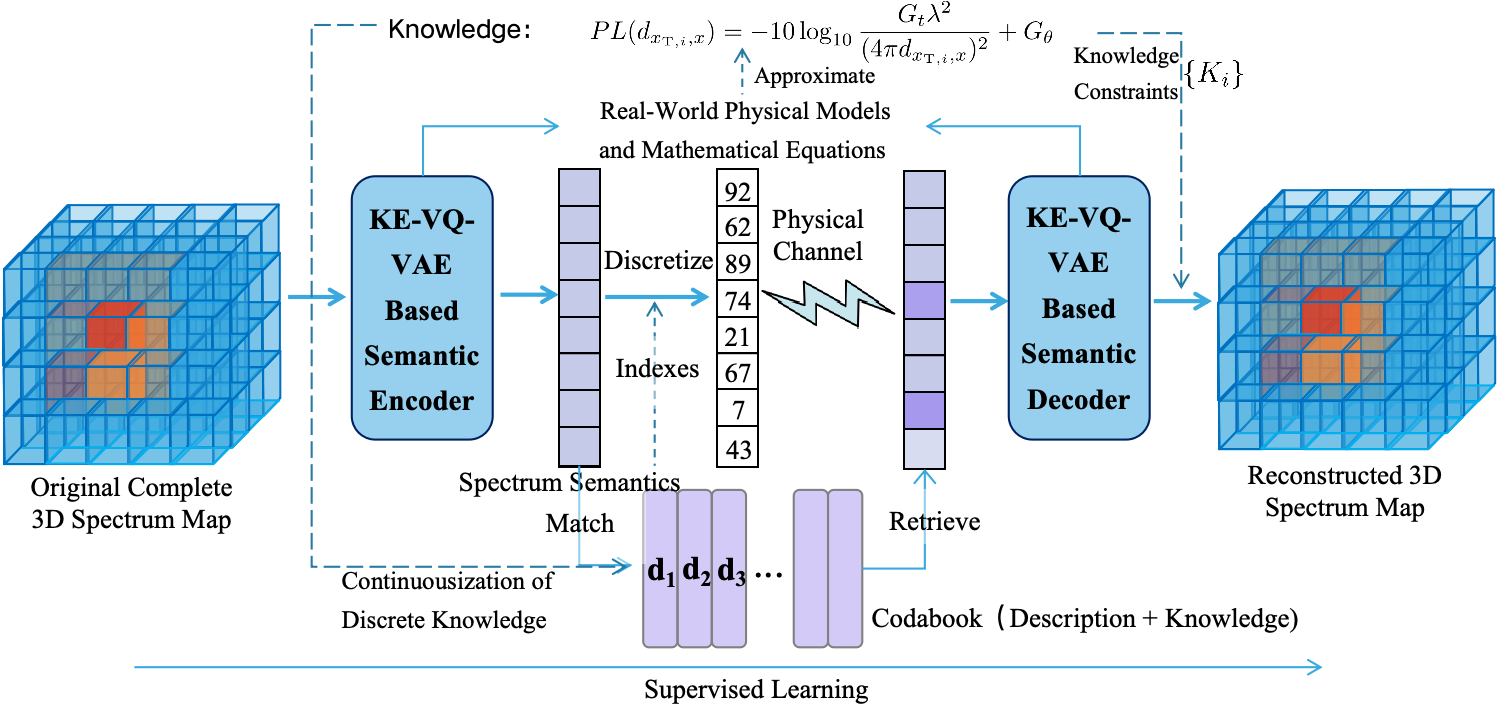} \\
	  \end{minipage}}
	\subfigure[Stage 2: The training of the 3D-Transformer based semantic predictor.]{
	  \begin{minipage}{15cm}
		\includegraphics[width=\textwidth]{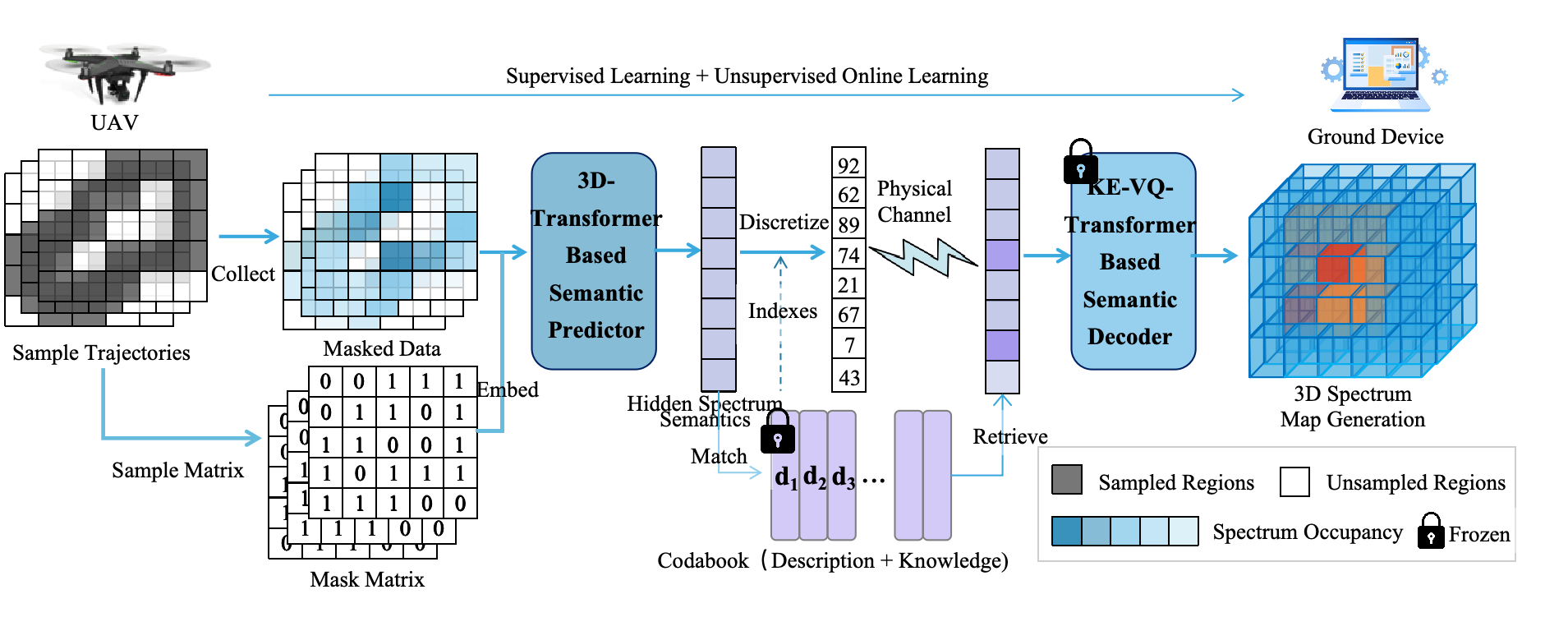} \\
	  \end{minipage}}
	\caption{Our proposed semantic spectrum map completion framework.} 
	\vspace{-0.3cm}
\end{figure*}

\subsection{Multi-Scale And Low-Complex 3D-Transformer}\label{lcd}
It is worth noting that we propose a knowledge-enhanced semantic communication framework for spectrum map reconstruction,
where the offline knowledge-enhanced semantic loss (\ref{offlinelossknow}) and online knowledge-enhanced semantic loss (\ref{onlinelossknow}) are implemented independently of the specific semantic coding network, i.e., they are applicable to any semantic communication network.
We note that 3D-Transformer can serialize spatial data and use self-attention mechanisms to capture long-range dependencies in the sequences,
which enables a better understanding of the correlations between different regions in the spatial spectrum map instead of being constrained by a fixed-size sensory field compared to the convolutional neural network (CNN).
Hence, the 3D-Transformer is introduced as the backbone of the knowledge-enhanced spectrum map completion framework.
However, it is challenging for the UAV to afford the ultra-large computation complexity of the multi-attention mechanism in 3D-Transformer.
To solve the complexity challenge and further enhance the map completion performance, 
a novel low-complex and multi-scale KE-VQ-Transformer-based semantic coding approach is proposed, 
which can more efficiently reconstruct the spectrum map compared to the normal 3D Transformer approach and the 3D CNN approach as demonstrated in \textbf{Section \ref{Sim}} with low computation complexity.

\subsubsection{Low-Complex Multi-Head Attention With Sparse Window}
The multi-attention mechanism is able to extract richer and more comprehensive feature information \cite{liu2021swin, arnab2021vivit, yuan2022monocular}.
However, it is challenging for the multi-head attention structure to address the 3D features, i.e., spatial attention, due to the ultra-large computation complexity.
For the semantics of 3D spectrum maps in this paper, it requires $N_{\mathrm{L}} \times N_{\mathrm{W}} \times N_{\mathrm{H}} \times N_{\mathrm{L}} \times N_{\mathrm{W}} \times N_{\mathrm{H}}$ attentions,
which are hard to calculate for UAVs with limited computation resources.
Note that one transmitter only affects the signal strength in the adjacent area,
we introduce the sparse window proposed in \cite{yuan2022monocular} for feasible 3D attention calculations with fewer spatial feature loss.

Firstly, the UAV measures the non-empty spectrum points $\{x_{\mathrm{M}, j}\}$ with the power features $\boldsymbol{\Omega}_f(\{x_{\mathrm{M}, j}\})$ in the map space.
Then we search all non-empty features within a $n \times n \times n$ sparse window centered on this feature and get the neighbor features 
$\left\{\boldsymbol{\Omega}_f(x_{\mathrm{N},j}), \forall x_{\mathrm{N},j} \in \Theta(x_{\mathrm{M},j})\right\}$,
where $\Theta(\cdot)$ represents the location set of adjacent features. 
Then the query, key, and value embeddings are calculated respectively by
$
Q_i=\mathcal{L}_Q(V_i), K_j=\mathcal{L}_K(V_j), V_j=\mathcal{L}_V(v_j),
$
where $\mathcal{L}_Q, \mathcal{L}_K, \mathcal{L}_V$ are the linear projection layers, and $v_j$ is the input value.
Hence, the 3D attention is calculated as
\begin{equation}
\mho(x_{\mathrm{M}, j})=\sum_{x_{\mathrm{N},j} \in \Theta(x_{\mathrm{M}, j})} \operatorname{softmax}(Q_{x_{\mathrm{M}, j}} K^{T}_{x_{\mathrm{N},j}} / \sqrt{C_k})(V_{x_{\mathrm{N},j}}),
\end{equation}
where $\mho(\cdot)$ is the attention function and $C_k$ represents the dimensionality of $K_{x_{\mathrm{N},j}}$.
In this case, the computation complexity of 3D attention is reduced
to $N_{\mathrm{NEF}} \times N_{\mathrm{NEN}}$,
where $N_{\mathrm{NEF}} = \Im(\{x_{\mathrm{M}, j}\})$ and $N_{\mathrm{NEN}} = \Im(\Theta(\{x_{\mathrm{M}, j}\}))$ are respectively the number of non-empty features and non-empty neighbor features within a local window size $N_{\mathrm{win}}$, 
where $\Im(\cdot)$ computes the number of elements in the set $(\cdot)$.

\subsubsection{Multi-Scale Spatial Spectrum Feature Extraction and Fusion}
Due to the different transmit strengths of the transmitters and the superposition of the signal strengths at the block, 
an irregular-like signal coverage can emerge in the spectrum map, which brings about the problem of scale variance.
Therefore, we adopt centralized-feature-pyramid based multi-scale spectrum space features with $R$ levels of scale, 
which can obtain the feature information under different receptive fields and improve the performance and accuracy of 3D spectral map reconstruction. 
For the $r$-th scale, $1 \leq r \leq R$, the resolution is $2^{1-r}$ times that of the original map.
Specifically, the high-level semantic coding network has a larger receptive field and a stronger ability to characterize the spectrum information, 
but it is difficult to characterize the geometric information, i.e., signal localization. 
The low-level network has a smaller receptive field, 
and the detailed geometric information characterization can complement the ambiguity of signal location in the high-level network.

\subsection{Stage 1: Training KE-VQ-Transformer Based Semantic Decoding}
In Stage 1, we aim to obtain the knowledge-enhanced multi-scale codebooks $\{\mathcal{D}_r\}$ and semantic decoders $\{E_{\beta_r}\}$, as shown in Fig. 2(a).
Specifically, at the simulated transmitter, the original completion 3D spectrum map $\boldsymbol{\Omega}_f$ is fed into the multi-scale
KE-VQ-Transformer based semantic encoder $\{\mathring{E}_{\mathring{\alpha}_r}\}$ to obtain the semantic features $\{\mathring{d}_{r}\}$ with different resolutions.
Based on  (\ref{ns}), the nearest semantic vectors $\{\widetilde{\boldsymbol{d}}_r\}$ are found in the codebooks, 
and the corresponding indexes of $\{\widetilde{\boldsymbol{d}}_r\}$ are transmitted over the simulation air-to-ground channel.
The ground user receives the indexes and recovers the semantics $\{\widehat{\boldsymbol{d}}_r\}$ by using the codebooks.
To construct the original spectrum map, the multi-scale map reconstruction proceeds step by step from low resolution to high resolution.
Specifically, the map features $\boldsymbol{\widehat{P}}_{R}$ with the $R$ scale can be obtained by 
\begin{equation}
	\boldsymbol{\widehat{P}}_{R} = E_{\beta_{R}}(\widehat{\boldsymbol{d}}_{R}).
\end{equation}
Then, the features $\boldsymbol{\widehat{P}}_{R}$ are upsampled to match the semantics $\widehat{\boldsymbol{d}}_{R-1}$ at the scale $R-1$.
Let the upsampled features be $\widetilde{\boldsymbol{P}}_{R}$, and the features are combined with the semantics at the channel dimension,
represented by $\widetilde{\boldsymbol{P}}_{R} \oplus \widehat{\boldsymbol{d}}_{R-1}$.
Hence, for the scale $1 \leq r \leq R-1$, the map features are iteratively generated and fusion,
represented by 
$
	\boldsymbol{\widehat{P}}_{r} = E_{\beta_{r}}(\widetilde{\boldsymbol{P}}_{R} \oplus \widehat{\boldsymbol{d}}_{R-1}), 1 \leq r \leq R-1.
$
Finally, the reconstructed map $\widehat{\boldsymbol{\Omega}}_f = \boldsymbol{\widehat{P}}_{1}$.

As the original completed maps are used to train the KE-VQ-Transformer-based semantic coding networks and the knowledge-enhanced codebooks,
Stage 1 is trained under offline learning, where the loss function is given in eq. (\ref{Off}).
The detailed information of the supervised training in Stage 1 is given in \textbf{Algorithm 1}.

\begin{algorithm}[t!]
	\caption{Offline Training of The Knowledge-Driven Multi- Scale Semantic Decoders And Codebooks In Stage 1.}
	\begin{algorithmic}[1]
	  \STATE Initialize the multi-scale semantic encoders $\{\mathring{E}_{\mathring{\alpha}_r}\}$, semantic decoders $\{D_{\beta_r}\}$, and codebooks $\{\mathcal{D}\}$;
	  \STATE \textbf{for} Epoch $m=1$ $\rightarrow$ $M$ \textbf{do}
	  
	  \textbf{The Simulated Transmitter:}
	  \STATE \par \hspace{0.8cm} Extract multi-scale features: $E_{\mathring{\alpha}}(\boldsymbol{\Omega}_f) \rightarrow \{\mathring{\boldsymbol{d}}_r\}_{r=1}^{R}$;
	  \STATE \par \hspace{0.8cm} Describe features:$\{\mathring{\boldsymbol{d}}_r \} \rightarrow \{\widetilde{\boldsymbol{d}}_r\}$, $\widetilde{\boldsymbol{d}}_r \in \mathcal{D}_r$;
	  
	  \textbf{The Simulated Air-to-Ground Channel:}
	  \STATE \par \hspace{0.8cm} Carry the indexes of $\{\widetilde{\boldsymbol{d}}_r\}$ over bits;
	  
	  \textbf{The Ground User:}
	  \STATE \par \hspace{0.8cm} Recover semantics: $\{\widehat{\boldsymbol{d}}_r\}_{r=1}^{R}$, $\widehat{\boldsymbol{d}}_r \in \mathcal{D}_r$;
	  \STATE \par \hspace{0.8cm} Generate map features: $\widehat{\boldsymbol{d}}_R \rightarrow \widehat{\boldsymbol{P}}_R$;
	  \STATE \par \hspace{0.8cm} \textbf{for} $r = R \rightarrow 1$ \textbf{do}
	  \STATE \par \hspace{1.6cm} Feature upsampling:  $\widehat{\boldsymbol{P}}_r \rightarrow \widetilde{\boldsymbol{P}}_r$;
	  \STATE \par \hspace{1.6cm} Fuse scale:  $E_{\beta_{r}}(\widetilde{\mathrm{P}}_{R} \oplus \widehat{\boldsymbol{d}}_{R-1}) \rightarrow \boldsymbol{\widehat{P}}_{r}$;
	  \STATE \par \hspace{0.8cm} \textbf{end for}
	  \STATE \par \hspace{0.8cm} Obtain reconstructed map: $\widehat{\boldsymbol{\Omega}}_f = \boldsymbol{\widehat{P}}_{1}$;
	  \STATE \par \hspace{0.8cm} Update $\{\mathring{\alpha}_r\}$, $\{\beta_r\}$ and $\{\mathcal{D}_r\}$ with the loss $\mathcal{L}^{\mathrm{Off}}$;
	  \STATE \textbf{end for}
	  \RETURN $\{\mathring{\alpha}^*_r\}$, $\{\beta_r^*\}$, $\{\mathcal{D}_r^*\}$.
	\end{algorithmic}
\end{algorithm}

\subsection{Stage 2: Training 3D-Transformer Based Semantic Prediction}
In Stage 2, the trained knowledge-enhanced semantic decoders $\{E_{\beta_r^*}\}$ and codebooks $\{\mathcal{D}_r^*\}$ are given,
and we aim to train the semantic predictors $\{E_{\alpha_r}\}$ that can extract semantics from measured spectrum points and predict the unknown information, as shown in Fig. 2(b).
Assume the semantic predictions are represented by $E_{\alpha_r}(\widetilde{\boldsymbol{\Omega}}_f)= \boldsymbol{d}_r$.
To enable $\{E_{\alpha_r}\}$ with the prediction ability, the semantic encoders $\{\mathring{E}_{\mathring{\alpha}^*_r}\}$ serve as the  ``teachers''.
Specifically, the description of the prediction results should be aligned with that of the features extracted by $\{\mathring{E}_{\mathring{\alpha}^*_r}\}$,
which can be represented by
\begin{equation}\label{argeq}
	f_r : \arg \min _{\boldsymbol{d}_{r,l_r}}\left\| \boldsymbol{d}_r - \widetilde{\boldsymbol{d}}_{r,l_r}\right\|_2 = \arg \min _{\boldsymbol{d}_{r,l_r}}\left\| \mathring{\boldsymbol{d}}_r - \widetilde{\boldsymbol{d}}_{r,l_r}\right\|_2 .
\end{equation}
In fact, (\ref{argeq}) can be regarded as a classification task, 
where the predictors can directly obtain the semantic symbol with probabilistic prediction.
Hence, (\ref{argeq}) can be rewritten as
\begin{equation}
	f_r : \arg \max_{\boldsymbol{d}_{r,l_r}} \boldsymbol{d}_r(\boldsymbol{d}_{r,l_r}) = \arg \min _{\boldsymbol{d}_{r,l_r}}\left\| \mathring{\boldsymbol{d}}_r - \widetilde{\boldsymbol{d}}_{r,l_r}\right\|_2.
\end{equation}
Here, the semantics $\boldsymbol{d}_r$ is the probability distribution of predictions with the size $L_r$,
and $\boldsymbol{d}_r(\boldsymbol{d}_{r,l_r})$ represents the probability of matching the $\boldsymbol{d}_{r,l_r}$.
We adopt the cross entropy loss to train the semantic predictors, represented by
\begin{equation}\label{cro}
	\mathcal{L}^{\mathrm{Cro}}=-\sum_{r=1}^{R}\sum_{l_r=0}^{L_r} y_{r,l_r} \log \left(\boldsymbol{d}_r(\boldsymbol{d}_{r,l_r})\right),
\end{equation}
where $y_{r,l_r}$ is the indicator of the $l_r$ classification at the scale $r$.
If the case $f_r$ happens, $y_{r,l_r} = 1$; otherwise, $y_{r,l_r} = 0$.

\begin{figure*}
	\centering
	\includegraphics[scale=0.8]{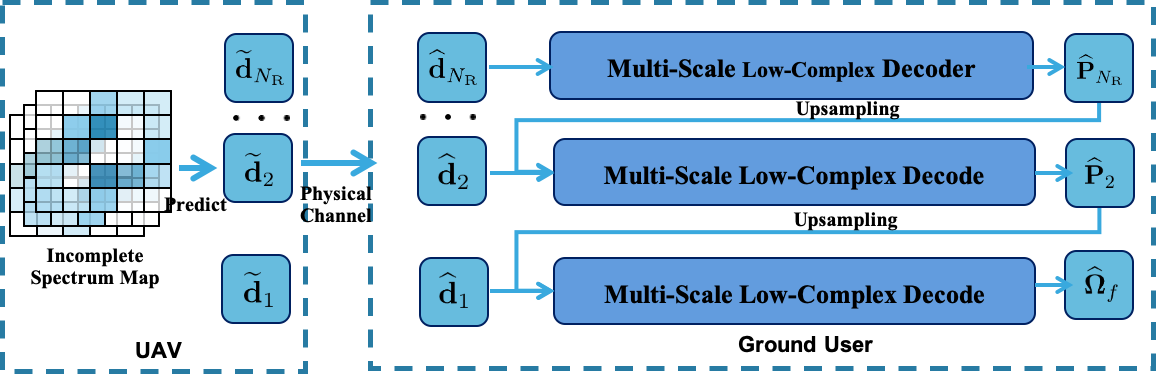}
	\vspace{-0.1cm}
	\caption{The overview of the semantic processing for 3D spectrum reconstruction.}
    \vspace{-0.2cm}
\end{figure*}
Further enhancing the completion performance in the practical dynamic environment, 
the whole network is tuned online by using the unsupervised loss $\mathcal{L}^{\mathrm{Onl}}$.
The details of joint offline training and online training on Stage 2 are presented in \textbf{Algorithm 2},
and the final processing is given in Fig. 3.

\subsection{Complexity Analysis Of The Proposed Algorithm}\label{ca}
Algorithm 1 and Algorithm 2 have the same computational complexity. 
Hence, we take Algorithm 1 as an example to analyze the complexity of the algorithms.
Let $C$ be the dimensionality of the maximum map patch, which equals the dimensionality of $\mathring{\boldsymbol{d}}_1$.
For the scale $r$, the complexity of computing $Q_i$, $K_j$ and $V_j$ is $\mathcal{O}(3N_{\mathrm{W}}N_{\mathrm{H}}N_{\mathrm{L}}(\frac{C}{2^{r-1}})^2)$.
Then, the computational complexity of multi-head attention is $\mathcal{O}(2(N_{\mathrm{NEF}} \times N_{\mathrm{NEN}})^2(\frac{C}{2^{r-1}}))$,
and the computational complexity of the projection is $\mathcal{O}(N_{\mathrm{W}}N_{\mathrm{H}}N_{\mathrm{L}}(\frac{C}{2^{r-1}})^2)$.
Thus, the total complexity is computed by $\sum_{r=1}^{R}\mathcal{O}(2(N_{\mathrm{W}}N_{\mathrm{H}}N_{\mathrm{L}})^2(\frac{C}{2^{r-1}})+4N_{\mathrm{W}}N_{\mathrm{H}}N_{\mathrm{L}}(\frac{C}{2^{r-1}})^2)
=\mathcal{O}(4\left(N_{\mathrm{NEF}} \times N_{\mathrm{NEN}}\right)^2 C\left(1-\frac{1}{2^R}\right)+\frac{16}{3} N_{\mathrm{W}} N_{\mathrm{H}} N_{\mathrm{L}} C^2\left(1-\frac{1}{4^R}\right))$. For the conventional full-window Transformer, the self-attention complexity grows quadratically with the total number of input tokens, i.e., $\mathcal{O}((N_{\mathrm{W}}N_{\mathrm{H}}N_{\mathrm{L}})^2)$.
In contrast, the proposed sparse-window mechanism divides the feature map into local windows of size $N_{\mathrm{win}} N_{\mathrm{win}} N_{\mathrm{win}}$, and the theoretical reduction ratio is $\frac{\mathcal{O}((N_{\mathrm{NEF}} \times N_{\mathrm{NEN}})^2C)}{\mathcal{O}((N_{\mathrm{W}}N_{\mathrm{H}}N_{\mathrm{L}})^2C)} 
< \frac{N_{\mathrm{win}} N_{\mathrm{win}} N_{\mathrm{win}}}{N_{\mathrm{W}}N_{\mathrm{H}}N_{\mathrm{L}}}.
$

\begin{algorithm}[t!]
	\caption{Joint Offline Training and Online Training of The Semantic Predictor and The Whole Network In Stage 2.}
	\begin{algorithmic}[1]
	  \STATE Given $\{\mathring{\alpha}^*_r\}$, $\{\beta_r^*\}$, $\{\mathcal{D}_r^*\}$;
	  \STATE Initialize the semantic predictors $\{E_{\alpha_r}\}$;
	  \STATE \textbf{for} Epoch $m=1$ $\rightarrow$ $M$ \textbf{do}
	  \STATE \par \hspace{0.8cm} Predict semantics: $E_{\alpha}(\widetilde{\boldsymbol{\Omega}}_f) \rightarrow \{\boldsymbol{d}_r\}_{r=1}^{R}$;
	  \STATE \par \hspace{0.8cm} Update the Predictors with the loss $\mathcal{L}^{\mathrm{Cro}}$;
	  \STATE \textbf{end for}
	  \RETURN $\{\alpha^*_r\}$

	  \textbf{The UAV:} \COMMENT{start online training}
	  \STATE Measure the signal power under constrainted trajectories: $\widetilde{\boldsymbol{\Omega}}_f$; 
	  \STATE Predict spatial features: $E_{\alpha^*}(\widetilde{\boldsymbol{\Omega}}_f) \rightarrow \{\widetilde{\boldsymbol{d}}_r\}_{r=1}^{R}$;
	  
	  \textbf{Wireless Channel:}
	  \STATE Carry indexes of $\{\widetilde{\boldsymbol{d}}_r\}$ over bits;
	  
	  \textbf{The Ground User:}
	  \STATE Recover semantics: $\{\widehat{\boldsymbol{d}}_r\}_{r=1}^{R}$, $\widehat{\boldsymbol{d}}_r \in \mathcal{D}_r$;
	  \STATE Generate map features: $\widehat{\boldsymbol{d}}_R \rightarrow \widehat{\boldsymbol{P}}_R$;
	  \STATE \textbf{for} $r = R \rightarrow 1$ \textbf{do}
	  \STATE \par \hspace{0.8cm} Feature upsampling:  $\widehat{\boldsymbol{P}}_r \rightarrow \widetilde{\boldsymbol{P}}_r$;
	  \STATE \par \hspace{0.8cm} Fuse scale:  $E_{\beta_{r}}(\widetilde{\mathrm{P}}_{R} \oplus \widehat{\boldsymbol{d}}_{R-1}) \rightarrow \boldsymbol{\widehat{P}}_{r}$;
	  \STATE \textbf{end for}
	  \STATE Obtain reconstructed map: $\widehat{\boldsymbol{\Omega}}_f = \boldsymbol{\widehat{P}}_{1}$;
	  \STATE Tine $\{\alpha^*_r\}$, $\{\beta^*_r\}$ and $\{\mathcal{D}^*_r\}$ by imposing unsupervised knowledge-driven constraints $\mathcal{L}^{\mathrm{Onl}}$ on $\widehat{\boldsymbol{\Omega}}_f$;	  
	  \RETURN $\{\bar{\alpha}^*_r\}$, $\{\bar{\beta}^*_r\}$ and $\{\bar{\mathcal{D}}^*_r\}$.
	\end{algorithmic}
\end{algorithm}

\section{Simulation Results}\label{Sim}
\subsection{System Parameter Settings}

\subsubsection{Spectrum Map Settings}
We consider a 3D spectrum space $\boldsymbol{S}$ with the dimension $L \times W \times H$,
where $L$, $W$ and $H$ are respectively set to $160$, $160$ and $120$ with the unit $\mathrm{m}$.
To evaluate the measurements, the space $\boldsymbol{S}$ is divided into $N_{\mathrm{L}} \times N_{\mathrm{W}} \times N_{\mathrm{H}}=64 \times 64 \times 24$ grids. 
Moreover, a training set with 30000 incomplete 3D radio maps and a testing set with 10000 are considered, and the dataset is created based on the open spectrum map generation source\footnote{https://github.com/fachu000/deep-autoencoders-cartography} provided by \cite{9523765}. 
For each 3D radio map, the transmitters are randomly placed on the ground (i.e., in the lowest $32 \times 32 \times 24$ grids), 
and the transmit power in the unit of $\mathrm{dBm}$ is randomly selected from the vector $[26,28,30,30,28,26]$. 
If there are no additional conditions, 
the number $N_{\mathrm{T}}$ of transmitters corresponding to the training set is randomly selected from 1 to 3, i.e., $N_{\mathrm{T}} \in\{1,2,3\}$. 
The frequency point $f$ of each map is set to $75~\mathrm{MHz}$, and the sampling ratio of the UAV is set to $\tau = 0.15$ if not otherwise specified.

\subsubsection{Computational Platform and Hardware Configuration}
All simulations and model training were carried out on a workstation equipped with an Intel Core i9-13900K CPU, 64 GB RAM, and a single NVIDIA RTX 4070 Super GPU with 12 GB VRAM. The model training required approximately 6 hours for convergence on the dataset. For on-board UAV evaluation, the trained model was deployed on an NVIDIA Jetson Orin~NX with 16\,GB LPDDR5 and 25\,W power mode.

\subsubsection{Channel model settings}
Let the channel between the UAV and the ground device be $\mathrm{h}^{U \rightarrow G}$, 
which is assumed to be the LoS path since the ground device is independent of the target spectrum area collected by the UAV.
Hence, the channel model $\mathrm{h}^{U \rightarrow G}$ is represented by
\begin{equation}
 \mathrm{h}^{U \rightarrow G}= \varpi d_0^{-\upsilon},
\end{equation}
where the $d$ denotes the space distance between the UAV and the ground device, 
$\varpi$ represents the channel gain with the reference distance $d^{\mathrm{ref}}=1 \mathrm{~m}$,
and $\upsilon$ represents the path loss exponent of the LoS path. 
Without loss of generality, for each spectrum semantic transmission, the distance $d_0$ is obtained randomly from the range [50,500] with the unit $m$.

\subsubsection{Semantic communication network settings}
We use the vision Transformer (ViT) \cite{dosovitskiy2020image} as the backbone of the features extractor/recoverer.
The number of semantic vectors in the codebook is set to $256$, 
where the index of the vectors is represented by $8$ bits.
The modulation method is set to 64QAM.
Moreover, the detailed semantic coding network is presented in \textbf{TABLE I}.

\begin{table*}
	\caption{Semantic Communication Network}
	\centering
	\begin{tabular}{|c|c|c|c|c|}
	  \hline
	  {}&Function& Layer & Units & Activation\\
	  \hline
	  \multirow{4}*{Simulated Transmitter}& \multirow{2}*{Semantic Encoder} &Patch Embedding & 64 & None\\
	  \cline{3-5}
	  & & 4 $\times$ 3D-Transformer & 64 & None\\ 
	  \cline{2-5}
	  & Vector Shift &Dense &64& Linear\\
	  \cline{2-5}
	  & Vector Quantization &Matching &64& None\\
	  \hline
	  \multirow{1}*{UAV}& Semantic Predictor &4 $\times$ 3D-Transformer & 64 & Softmax\\
	  \hline
	  Wireless Channel& Channel& None &None &None\\
	  \hline
	  \multirow{3}*{Receiver}&Codebook&Mapping& 64 &None\\
	  \cline{2-5}
	  &\multirow{2}*{Semantic Decoder}&4$\times$ 3D-Transformer & 64 &None\\
	  \cline{3-5} 
	  & & Patch Debedding & Map Size & None\\ 
	  \hline
	\end{tabular}
	\vspace{-0.3cm}
\end{table*}

\subsubsection{Baseline scheme settings}
The inverse distance weighted (IDW) is introduced as a mathematical spatial interpolation method, which can estimate the unknown signal power 
based on the signal power and positions of known spectrum points.
The IDW formula for 3D spectrum map completion can be written as 
\begin{equation}
	Z(x, y, z)=\frac{\sum_{i=1}^{N_{\mathrm{S}}} \frac{Z_i}{\frac{1}{p}}}{\sum_{i=1}^{N_{\mathrm{S}}} \frac{1}{d_i^p}},
\end{equation}
where $x$, $y$ and $z$ represent the 3D coordinates of the spectrum map,
$Z_i$ represents the signal power of the $i$-th known sample point,
$d_i$ represents the Euclidean distance between the target spectrum point $(x, y, z)$ and the $i$-th known spectrum point ($x_i$, $y_i$, $z_i$),
$p$ is the factor controlling the influence of distance,
and $N_{\mathrm{S}}$ represents the number of the sampled spectrum points.
The Euclidean distance $d_i$ is obtained by
$d_i=\sqrt{\left(x-x_i\right)^2+\left(y-y_i\right)^2+\left(z-z_i\right)^2}$.
To recover the complete map at the ground user with IDW, the original sampled data is transmitted over the channel by LDPC+64QAM. 

We extend the two dimensional (2D)-CNN based deep autoencoder (CNN-DAE) scheme \cite{9523765} 
into the 3D spectrum space. Specifically, the 2D-CNN architecture is replaced by the 3D-CNN architecture,
and the codebook-based semantic discretization approach, the same as our proposed scheme, is used over the wireless channel. Moreover, the GAN-based spectrum map completion (GAN-SMC) scheme \cite{10025551} is introduced.
As the backbone network is replaced by a 3D-Transformer in CNN-DAE, the Transformer-DAE is also considered.
To further explore the impact of the knowledge-driven constraints and the multi-scale framework,
the VQ-Transformer scheme without knowledge-driven constraints and the single-scale KE-VQ-Transformer scheme are considered.

\subsection{Convergence Performance Analysis}
\begin{figure}
	\centering
	\includegraphics[scale=0.6]{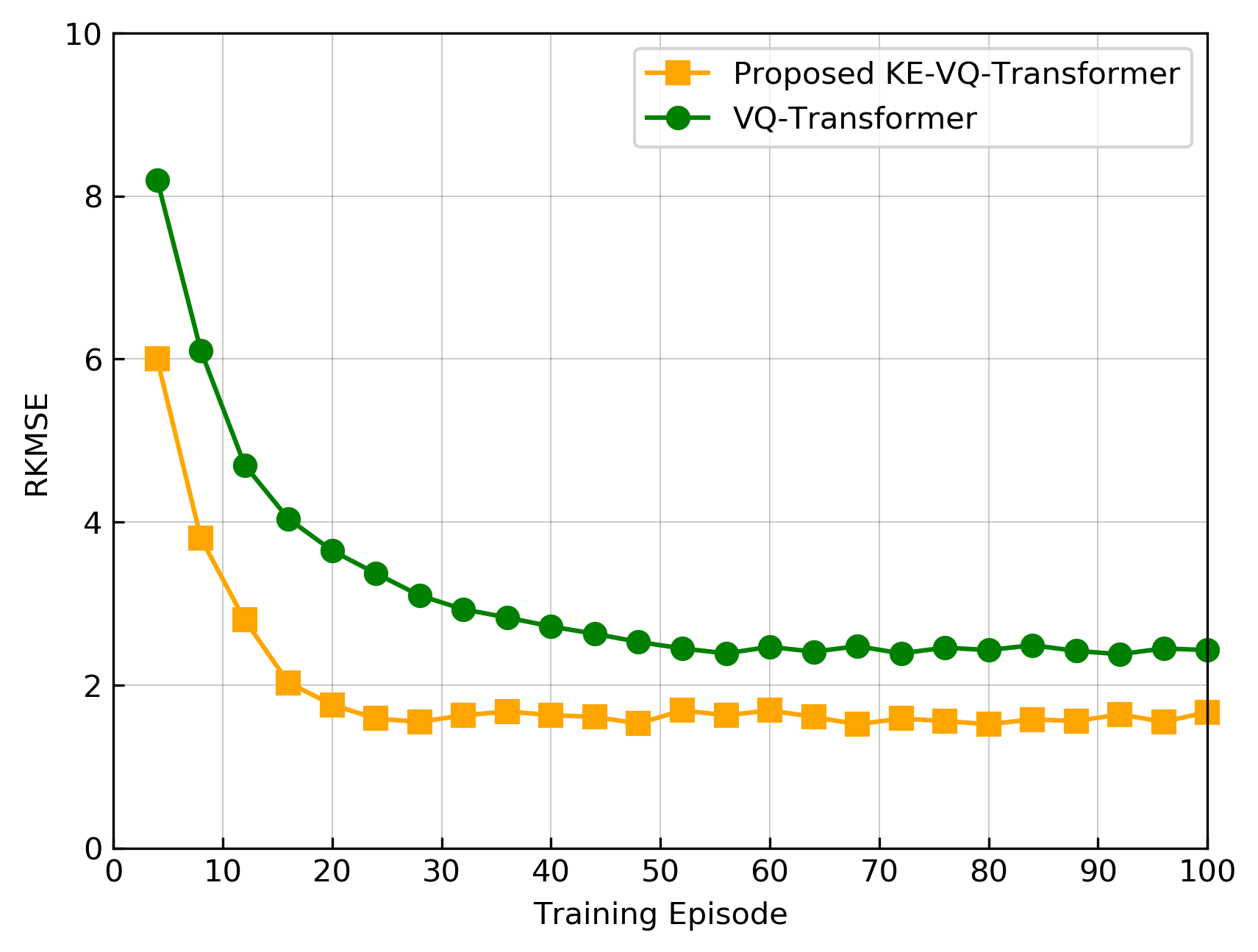}
	\vspace{-0.3cm}
	\caption{The convergence performance of Stage 1 when SNR = 12 $\mathrm{dB}$ and $\tau = 0.15$.}
	\vspace{-0.3cm}
\end{figure}

\begin{figure}[t!]
	\centering
	\includegraphics[scale=0.6]{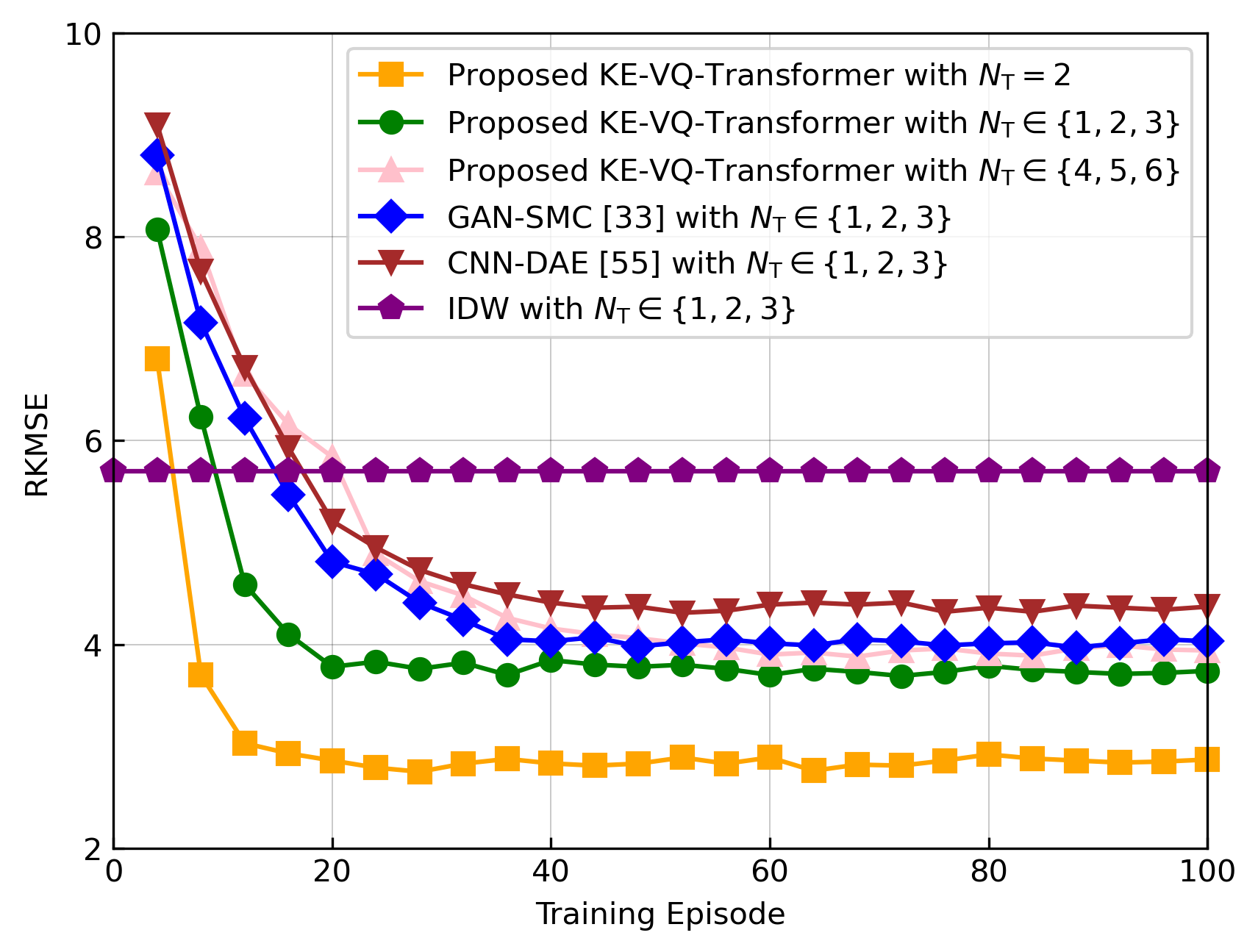}
	\vspace{-0.3cm}
	\caption{The convergence performance of Stage 2 with $\tau = 0.15$.}
	\vspace{-0.1cm}
\end{figure}
Fig. 4 demonstrates the convergence of Stage 1, where the SNR = 12 $\mathrm{dB}$ and $\tau = 0.15$.
It is seen that the KE-VQ-Transformer achieves convergence around 25 episodes while the VQ-Transformer 
achieves convergence around 50 episodes. 
This means even with the completed spectrum maps, the DL-based semantic spectrum map reconstruction can enable more realistic modeling of the physical world by introducing our proposed knowledge function.
Fig. 5 further illustrates the convergence of Stage 2 with dynamic numbers of the transmitters,
where the GAN-SMC scheme, CNN-DAE scheme and the IDW scheme are introduced as the baselines.
In Stage 2, the achievable convergence of the KE-VQ-Transformer, GAN-SMC, CNN-DAE, and the IDW is, respectively, around 10, 40, 20, and 40 episodes.
The convergence speed and the RKMSE performance of the proposed scheme demonstrate the effectiveness of the proposed sparse window and multi-scale design.
It can be seen that our proposed schemes outperform the baseline schemes in terms of the convergence speed and RKMSE. 
This is due to the fact that the knowledge-driven constraints imposed on the KE-VQ-Transformer
can effectively guide the model to fit a real physical world model. Moreover, with a larger number of transmitters, $N_{\rm{T}}=\{4,5,6\}$, the proposed model still achieves consistent convergence. 
This demonstrates that the embedded spectrum-knowledge constraints effectively regularize the learning process even when overlapping regions occur in high-density transmitter deployments.
Fig. 6 investigates the necessity of the proposed online training.
We can observe that the online training scheme achieves a 20\% improvement in terms of RKMSE compared to the sole offline training scheme.
This is because the online knowledge loss enables the semantic model
aware of the free space signal propagation model without supervision in a dynamic environment.

\subsection{Computational Complexity}
Table~II shows the comparisons of computational complexity in terms of floating point operations (FLOPs), memory usage, and inference time over different baselines. 
Although the theoretical complexity of the proposed Transformer is $\mathcal{O}(4(N_{\mathrm{NEF}}\!\times\!N_{\mathrm{NEN}})^2C)$, the sparse-window attention and hierarchical patching significantly reduce the number of active tokens at each scale. 
Consequently, the total computational cost is reduced by 25\% compared with a conventional VQ-Transformer. 
Moreover, the proposed KE-VQ-Transformer achieves 72\,G FLOPs and 2.08\,GB memory consumption, which requires only 53.9\,ms inference latency on the Jetson Orin NX and 26.4\,ms on the RTX 4070 Super. 
In contrast, the GAN-SMC scheme demands 110\,G FLOPs and 2.81\,GB memory. It indicates that the proposed knowledge-enhanced sparse design performs high reconstruction accuracy while providing substantial reductions in both computation and memory usage.

\begin{figure}[t!]
	\centering
	\includegraphics[scale=0.6]{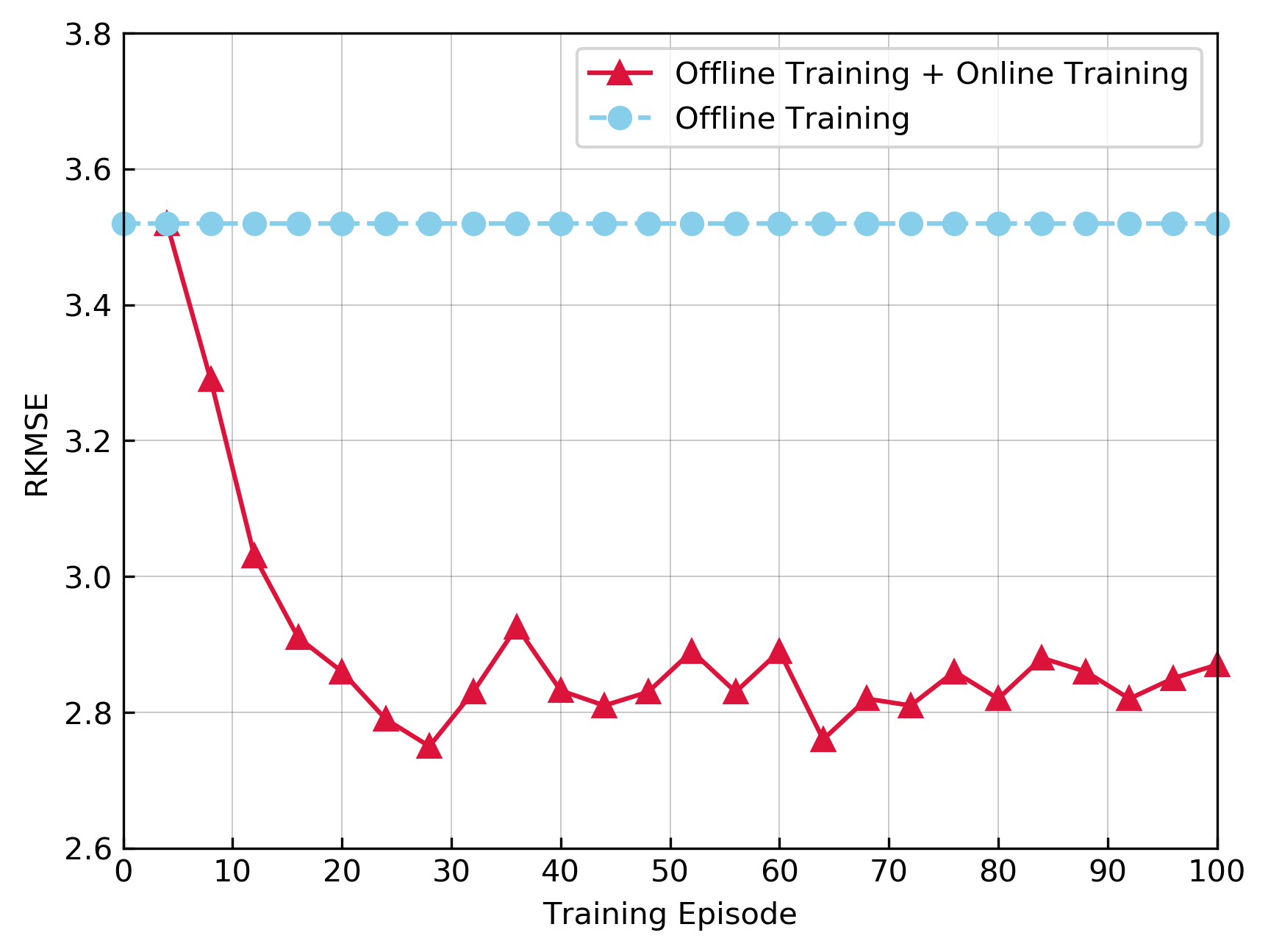}
	\vspace{-0.7cm}
	\caption{The completion performance of single offline training and joint offline and online training.}
	\vspace{-0.3cm}
\end{figure}

\begin{table}[t!]
\centering
\caption{Comparisons of computational complexity in terms of FLOPs, memory usage, and inference time over different baselines.}
\begin{tabular}{lcccc}
\toprule
\multirow{2}{*}{Model} &
\multirow{2}{*}{FLOPs (G)} &
\multirow{2}{*}{Mem. (GB)} &
\multicolumn{2}{c}{Inference latency (ms)} \\
\cmidrule(lr){4-5}
 & & & UAV & Ground \\
\midrule
IDW (GPU) & 4.8 & 0.42 & {\textbackslash} & 9.2 \\
VQ-Transformer & 95 & 2.36 & 78.9 & 30.1 \\
GAN\text{-}SMC~\cite{10025551} & 110 & 2.81 & 97.6 & 35.8 \\
CNN\text{-}DAE~\cite{9523765} & 32 & 1.15 & 43.8 & 18.9 \\
\textbf{KE-VQ (Ours)} & \textbf{72} & \textbf{2.08} & \textbf{53.9} & \textbf{26.4} \\
\bottomrule
\end{tabular}
\end{table}

Table~III shows performance of the proposed KE-VQ-Transformer under different window sizes $N_{\mathrm{win}}$. 
Increasing $N_{\mathrm{win}}$ enlarges the receptive field of each attention block, allowing the model to capture broader semantic correlations within the 3D spectrum map, thus enhancing the map reconstruction accuracy. 
It can be observed that the performance improvement saturates when $N_{\mathrm{win}}>8$, while the computational overhead continues to grow. 
This balance demonstrates that the sparse-window mechanism effectively limits redundant global computation while preserving sufficient contextual information for semantic spectrum-map reconstruction.

\begin{table}[t!]
\centering
\caption{Ablation on window size $N_{\mathrm{win}}$ for KE-VQ-Transformer.}
\begin{tabular}{cccc}
\toprule
$N_{\mathrm{win}}$ & KMSE  & FLOPs (G) & Inference (ms) \\
\midrule
4 & 4.128  & 48 & 42.3 \\
6 & 3.609 & 58 & 47.1 \\
\textbf{8} & \textbf{3.103}  & \textbf{72} & \textbf{53.9} \\
10 & 3.159 & 86 & 61.2 \\
\bottomrule
\end{tabular}
\vspace{-0.3cm}
\end{table}

\subsection{Completion Performance Analysis}
\begin{figure*}
	\centering
	\includegraphics[scale=0.35]{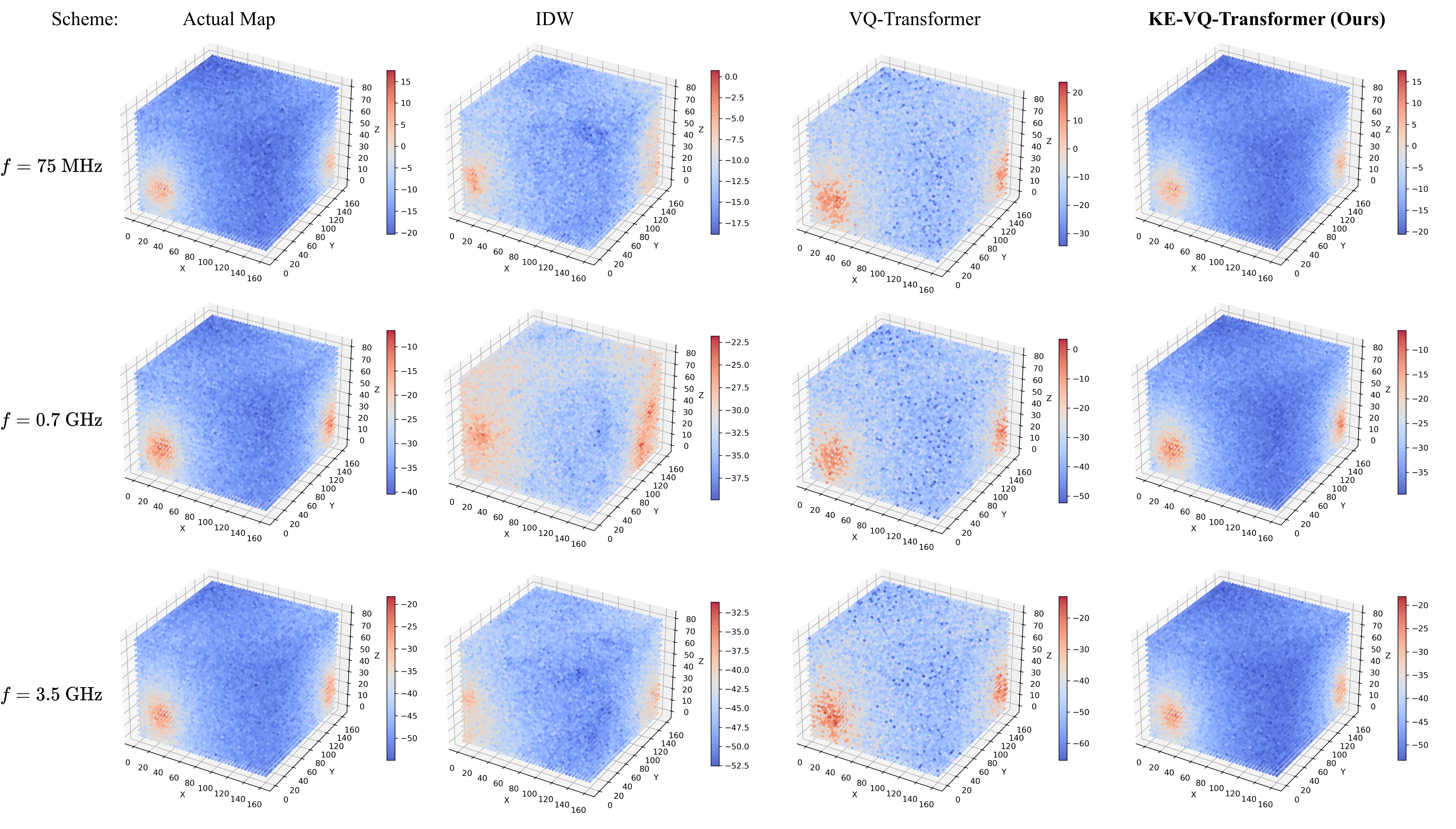}
	\vspace{-0.3cm}
	\caption{The visual map completion performance of different schemes when $\tau = 0.15$, $N_{\mathrm{T}}=4$ and the transmit power = 30 $\mathrm{dBm}$.}
	\vspace{-0.35cm}
\end{figure*}
Fig.~7 illustrates the visualization results of different schemes under a complex multi-transmitter scenario, where the number of transmitters is set to $N_{\mathrm{T}}=4$, the sampling rate is set to $\tau = 0.15$, and the operating frequencies are set to 75~MHz, 0.7~GHz for LTE low-band and 3.5~GHz for 5G FR1. 
The received signal power points are color-coded in the 3D spectrum map. 
It can be observed that the proposed KE-VQ-Transformer achieves more accurate reconstruction than both the VQ-Transformer and IDW. 
From the visualization maps, IDW and VQ-Transformer fail to preserve the physical propagation continuity and show blurred or distorted boundaries around transmitter regions. 
This degradation occurs because these data-driven baselines depend solely on statistical correlations among the observed samples, without embedding the underlying propagation knowledge. 
In contrast, the proposed KE-VQ-Transformer leverages spectrum-domain prior knowledge to maintain physical consistency across spatial scales, thereby achieving sharper boundary reconstruction and improved robustness under dense and realistic 4G/5G spectrum conditions.

\begin{figure}
	\centering
	\includegraphics[scale=0.6]{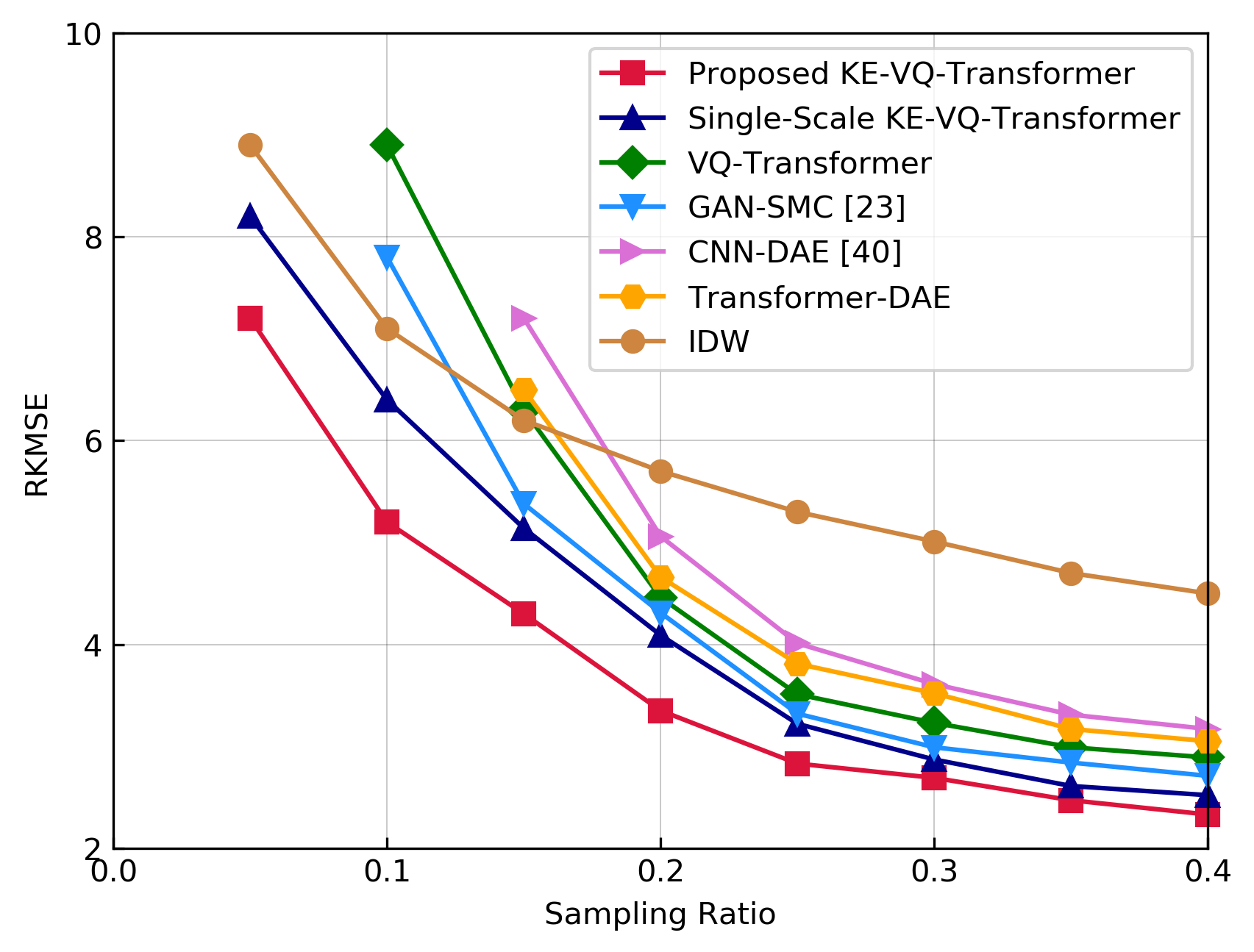}
	\vspace{-0.3cm}
	\caption{The map completion performance verus different sampling ratios with SNR = 12 $\mathrm{dB}$ in terms of the average RKMSE over 100 tests.}
	\vspace{-0.2cm}
\end{figure}
In Fig. 8, we investigate the impact of the sampling ratio on our proposed scheme and several benchmark schemes including VQ-Transformer, GAN-SMC, CNN-DAE, Transformer-DAE and IDW.
It can be seen that the knowledge-enhanced schemes perform better completion efficiency due to the refined knowledge-driven constraints.
Moreover, we can observe that the semantic spectrum completion schemes can outperform the traditional IDW scheme.
This demonstrates that our proposed AI-native semantic communication for spectrum maps can simultaneously reduce transmission overhead and improve complementary accuracy.
Note that, at the extremely small sampling ratio, the proposed KE-VQ-Transformer scheme significantly outperforms the benchmark schemes.
This is due to the fact that the proposed scheme can jointly leverage the data patterns and physical signal propagation knowledge, 
and the sampled points can serve as the progressive constraints on the map construction.
In contrast, the benchmark schemes directly utilized the known power points to complete the unknown power points,
which is thus heavily influenced by the number of sampling points.

\begin{figure}
	\centering
	\includegraphics[scale=0.6]{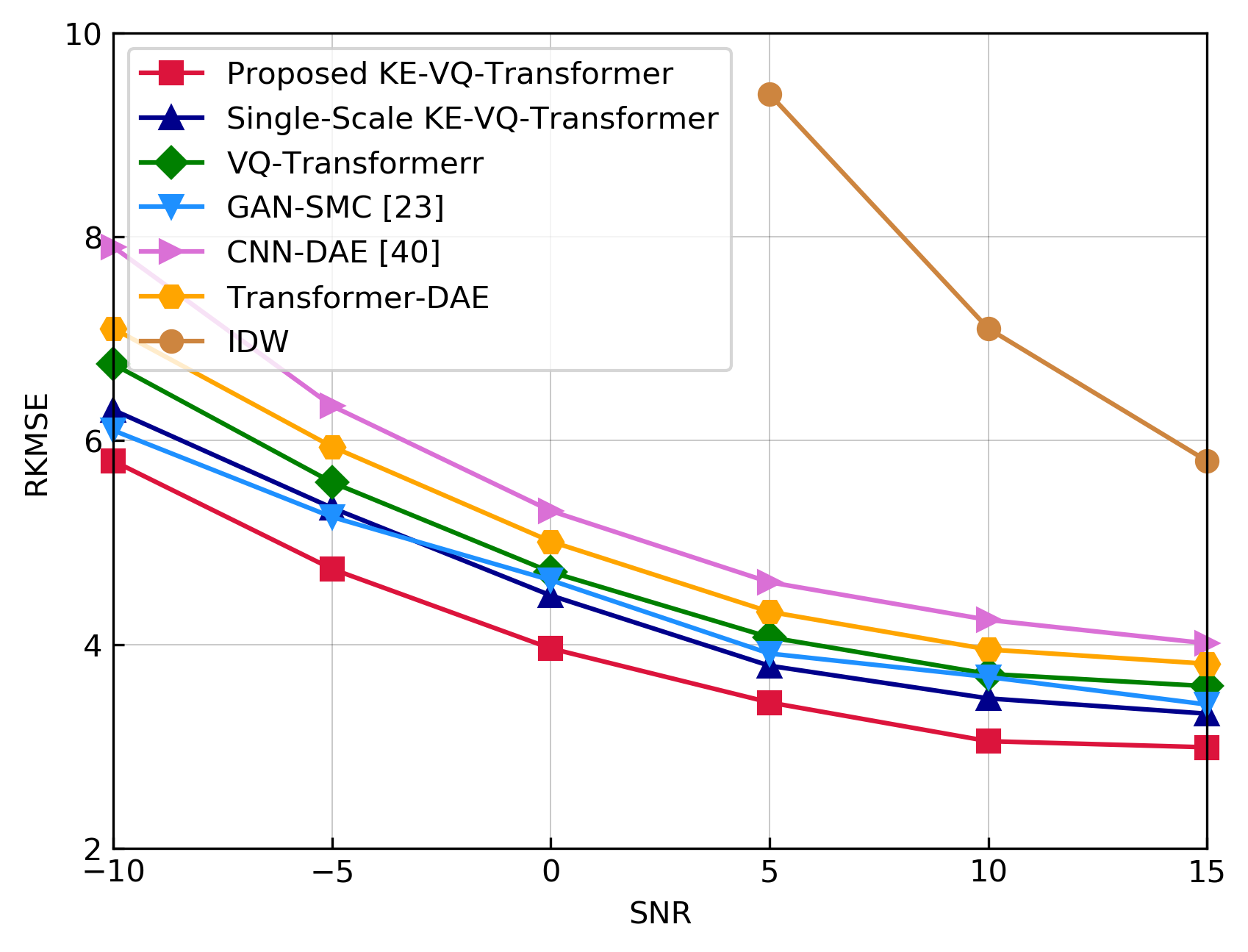}
	\vspace{-0.3cm}
	\caption{The map completion performance verus different SNR with $\tau = 0.2$ in terms of the average RKMSE over 100 tests.}
	\vspace{-0.5cm}
\end{figure}
In Fig. 9, we investigate the impact of the different SNRs on our proposed scheme and the same benchmark schemes with Fig. 8.
It is seen that our proposed scheme can achieve the lowest RKMSE compared to the benchmark schemes even at the low SNRs.
This is because our proposed knowledge-driven constraints can enable the generated spectrum map to follow the physical free space signal propagation model even with inaccurate semantics at low SNRs. 
With SNR = 0 $\mathrm{dB}$, our proposed KE-VQ Transformer can achieve up to 12\% performance improvement compared to the Single-Scale KE-VQ-Transformer and 13\% performance improvement compared to VQ-Transformer in terms of RKMSE,
which are respectively regarded as the gains from the multi-scale design and the knowledge-driven constraints.
It is worth noting that IDW is a typical scheme for map completion in traditional communication networks,
which performs weak completion performance or even failure at the low SNRs.
This further emphasizes the necessity of investigating semantic communication for intelligent spectrum map completion,
which enables robust map completion even at low SNRs.

\section{Conclusion}
Regarding the challenges of the missing semantic metric for spectrum map completion, the ultra-large computation overhead of 3D attention, and black-box machine learning, 
we proposed a novel knowledge-driven semantic spectrum completion framework for the first time.  
We extracted two expert prior knowledge-driven constraints from the free space signal propagation model to 
enable the semantic coding network be aware of the real-world physical model, thus enhancing the robustness and interpretability of semantic spectrum map completion.
Based on the knowledge-enhanced semantic spectrum map framework, we proposed a KE-VQ-Transformer based multi-scale low-complex semantic communication network, 
where the sparse window significantly reduced the 3D attention computation, and the multi-scale design further improved the completion performance for the 3D spectrum map.
The KMSE was defined for the semantic spectrum map completion to jointly consider the MSE and the physical deviation from the physical free space signal propagation model. 
A two-stage training method with joint offline and online training was conducted by imposing both supervised and unsupervised knowledge-driven constraints. 
The simulation results demonstrated that the proposed knowledge-driven scheme could learn from the physical signal propagation model and achieve superior completion performance in terms of RKMSE.
\bibliography{IEEEabrv,mybibfile}

\end{document}